\documentclass[twocolumn, twocolappendix]{aastex631}

\usepackage{amsmath}
\usepackage{xcolor}

\accepted{August 24, 2022}

\submitjournal{AJ}

\shorttitle{Low-mass Star Mass-Magnitude Relation}
\shortauthors{Giovinazzi et al.}

\begin{document}

\title{A Mass-Magnitude Relation for Low-mass Stars Based on \\
Dynamical Measurements of Thousands of Binary Star Systems }

\author[0000-0002-0078-5288]{Mark R. Giovinazzi}
\affiliation{Department of Physics and Astronomy, University of Pennsylvania \\
209 South 33rd Street, Philadelphia, PA 19104 USA}

\author[0000-0002-6096-1749]{Cullen H. Blake}
\affiliation{Department of Physics and Astronomy, University of Pennsylvania \\
209 South 33rd Street, Philadelphia, PA 19104 USA}

\begin{abstract}

Stellar mass is a fundamental parameter that is key to our understanding of stellar formation and evolution, as well as the characterization of nearby exoplanet companions. Historically, stellar masses have been derived from long-term observations of visual or spectroscopic binary star systems. While advances in high-resolution imaging have enabled observations of systems with shorter orbital periods, stellar mass measurements remain challenging, and relatively few have been precisely measured. We present a new statistical approach to measuring masses for populations of stars. Using Gaia astrometry, we analyze the relative orbital motion of $>3,800$ wide binary systems comprising low-mass stars to establish a Mass-Magnitude relation in the Gaia $G_\mathrm{RP}$ band spanning the absolute magnitude range $14.5>M_{G_\mathrm{RP}}>4.0$, corresponding to a mass range of $0.08$~M$_{\sun}\lesssim M\lesssim1.0$~M$_{\sun}$. This relation is directly applicable to $>30$ million stars in the Gaia catalog. Based on comparison to existing Mass-Magnitude relations calibrated for 2MASS $K_{s}$ magnitudes, we estimate that the internal precision of our mass estimates is $\sim$10$\%$. We use this relation to estimate masses for a volume-limited sample of $\sim$18,200 stars within 50~pc of the Sun and the present-day field mass function for stars with $M\lesssim 1.0$~M$_{\sun}$, which we find peaks at 0.16~M$_{\sun}$. We investigate a volume-limited sample of wide binary systems with early K dwarf primaries, complete for binary mass ratios $q>0.2$, and measure the distribution of $q$ at separations $>100$~au. We find that our distribution of $q$ is not uniformly distributed, rather decreasing towards $q=1.0$.

\end{abstract}


\section{Introduction} \label{sec:intro}

The measurement of stellar mass is important for addressing a wide range of scientific questions. The evolution of an isolated star is primarily determined by its initial stellar mass. While metallicity and angular momentum play a role, it is mass that dictates the path the star will take through the Hertzsprung-Russel (H-R) diagram as it evolves, the timescale of that evolution, and what will remain of the star after its fusion-powered lifetime. As one of the central predictions for star formation, measuring the Initial Mass Function is crucial to constraining those theories. The total stellar mass of the Milky Way and other galaxies is important for understanding the dynamics and evolution of the galactic environment. As the pace of exoplanet discovery has increased over the last decade, and the precision of the stellar measurements that enable the indirect detection of exoplanets has improved, precise knowledge of the host star mass is becoming an important limitation in efforts to characterize the bulk properties of these exoplanets. 

Despite all these compelling scientific reasons to measure stellar mass, even today relatively few stellar masses have been measured directly in a model-independent way. This is because stellar mass remains extremely difficult to measure, with no known observation of a single star that directly and precisely constrains mass (see \citealt{Serenelli2021} for a review of the methods of measuring stellar mass across the H-R diagram). Direct, model-independent stellar masses can be measured with observations of stellar binary systems. Both detached, double-lined eclipsing binary systems and fully resolved visual binaries combined with stellar radial velocity (RV) measurements can, in principle, produce individual stellar mass estimates with $2\%$ accuracy or better (\citealt{Serenelli2021}). Particularly when extremely precise RVs and astrometry from Gaia can be combined, impressive precision in mass measurements can be obtained (see, for example, \citealt{Brandt2019}). However, there are precious few cases where stellar mass is directly measured using dynamical techniques with accuracy at the $<2\%$ level. Fewer than 600 stars in 300 eclipsing binary systems meet this criteria today in the catalog described by \cite{Southworth2015}. The cases where stellar age and metallicity are well-estimated in addition to mass, the so-called benchmark systems that can be directly compared to stellar evolutionary models, are many fewer in number. Other, usually less precise, model-independent methods for estimating stellar mass have also been demonstrated, including the use of photometric ``flicker" to infer surface gravity mass (see, for example, \citealt{Stassun2018}), and in rare cases gravitational lensing effects (\citealt{Kluter202}).

Low-mass stars evolve relatively slowly, so it becomes possible to estimate mass directly from luminosity without degeneracies related to stellar evolution off the main sequence. For example, according to the MIST evolutionary tracks, a 0.7~M$_{\sun}$ star will change brightness in absolute Gaia $G_\mathrm{RP}$ magnitude by only 0.25 magnitudes between an age of 1-10~Gyr (\citealt{MIST2016}). A number of authors have developed empirical Mass-Magnitude relations that use precise, directly measured low-mass star masses to relate observed absolute magnitude to mass for stars on the lower main sequence (see, for example, \citealt{Delfosse2000}, \citealt{Mann_2019}, and references therein). Since infrared magnitudes are expected to be less sensitive to metallicity than optical magnitudes, \cite{Mann_2019} used precise dynamical mass measurements of 62 low-mass stars to develop a Mass-Magnitude relation using 2MASS $K_s$ magnitudes having an impressive internal precision of approximately $2\%$ across a mass range of $0.075<\rm{M_{\sun}}<0.7$. \citet{Mann_2019} demonstrated that this relation is insensitive to metallicity over the typical range of metallicities found in the solar neighborhood. However, the applicability of this relation is somewhat limited by the modest depth of 2MASS ($K_s<14.0$, in general) and the relatively small number of reference mass measurements used to calibrate the relation. 

The Gaia satellite provides photometric and astrometric measurements of tens of millions of low-mass stars \citep{Gaia2016, Gaia2021}. Statistically, many of these stars are expected to be in gravitationally bound systems (\citealt{Raghavan2010}), some of which will have projected separations large enough to be easily resolved by Gaia. In these cases, the projected \textit{relative} orbital motion of the system may be directly measured. Given the astrometric precision of Gaia (approximately 70~$\mu$as yr$^{-1}$ in proper motion at $G=17$), this relative orbital motion is well-measured for a large number of wide binary systems. The Gaia eDR3 release includes only linear components of the projected stellar motion, so it is not possible to solve for Keplerian orbits or directly estimate the total mass of the system with the current public Gaia data alone. However, given the large number of wide binary systems observed by Gaia, it does become possible to consider the overall properties of this relative orbital motion to directly constrain stellar mass in a statistical sense. A similar approach has been taken by \citet{Tokovinin2016} and \citet{Hwang2022} to constrain the overall eccentricity distribution of populations of binary stars.

Beginning with the catalog of Gaia binary systems assembled by \citet{Elbadry2021}, we carry out an analysis of relative orbital motion for 3,846 low-mass star systems to directly calibrate a modified linear Mass-Magnitude relation in the Gaia $G_\mathrm{RP}$ band that extends from the bottom of the main sequence up to 1.0~M$_{\sun}$ with typical errors of $<10\%$ on a per-object basis. In our analysis, we marginalize over the population of binaries that appear to have non-physical relative orbital motion under the assumption that the majority of our systems are gravitationally bound and contain no additional components beyond the binary pair. Our analysis is model-dependent only through an assumption about the underlying binary eccentricity distribution, which we find has a negligible impact on the resulting Mass-Magnitude relation. We apply our Mass-Magnitude relation to Gaia observations of 18,187 single low-mass stars to estimate the field mass function over a volume-limited sample. We also use a volume-limited sample of wide binaries having an early K-type primary star to estimate the mass ratio distribution $q$ and find that $q$ decreases towards 1.0 for our sample, which is complete for $q>0.2$. 

In Section \ref{sec:muprime}, we discuss the relative astrometric orbits of resolved binary star systems in the context of precise proper motion measurements from Gaia. In Sections \ref{sec:samp_select} and \ref{sec:measure}, we describe the selection of a sample of wide binary systems from the Gaia eDR3 database following the work of \citet{Elbadry2021}. In Section \ref{sec:LLH}, we detail the statistical framework we use to infer stellar mass within subsamples of the larger wide binary sample. In Section \ref{sec:results}, we compare our mass estimates to others in the literature and estimate the internal precision of our Mass-Magnitude relation. In Section \ref{sec:apps}, we apply the modified linear Mass-Magnitude relation to estimate the mass function of field stars with masses $<1.0~\mathrm{M}_{\odot}$ within 50~pc of the Sun and the mass ratio distribution of a sample of wide binaries.

\section{Relative Binary Orbits} \label{sec:muprime}

For visual binary systems, where the orbital motion of one star relative to the other can be measured, observations spanning a significant portion of the orbital period can yield a direct, model-independent estimate of the total mass of the system through Kepler's third law. Since the barycentric motion of the system as it orbits the Milky Way is usually not known independently, and absolute positions of the stars are more difficult to measure than relative positions, it is most often this relative orbit that is considered. In some cases, observations of relative orbits over even relatively short observational baselines can yield excellent mass estimates (e.g. \citealt{lucy2014}, \citealt{Brandt2019}, \citealt{Pearce_2020}). This fundamental dynamical technique has been used for more than a century and forms the basis of our understanding of stellar mass across the H-R diagram (e.g. \citealt{80Tau}).

A full astrometric description of a relative Keplerian orbit requires six free parameters, so a single measurement of the projected (on the sky) relative motion and separation of two stars in a binary system does not enable a unique solution for the system (see, for example, \citealt{Pearce_2020}). However, measurements of the relative proper motions of spatially resolved binaries can be used to constrain the likely orbital parameters of the system in a statistical sense. For example, \citet{tokovinin1998}, \citet{shatsky2001}, \citet{Tokovinin2016}, and \cite{Hwang2022}, investigated the angle of the relative orbital motion in a binary system, which is denoted $\gamma$, and found that the distribution of this angle can be used to infer the properties of the eccentricity distribution of the overall binary population, or generate broad posteriors on the eccentricities of individual systems.

In addition to the angle of relative motion, the projected speed of the relative orbital motion, which we denote $\mu$, can be measured as the difference in the proper motions of the two stars. For a circular, equal-mass, face-on binary system of mass $M_\mathrm{tot}$, semi-major axis $a$, and distance $d$, the relative orbital motion of the system is given by
\begin{equation}
    \mu = 2.0~\left(\frac{M_\mathrm{tot}}{\rm{M}_{\sun}}\frac{1000~\mathrm{au}}{a}\right)^{0.5}\left(\frac{100~\rm{pc}}{d}\right) \rm{mas}~\rm{yr}^{-1}.
\label{eqn:muscale}
\end{equation}
Given Gaia's reported uncertainties in proper motion, which are approximately $0.07$~mas~yr$^{-1}$ at $G=17$ and $0.5$~mas~yr$^{-1}$ at $G=20$ (\citealt{Gaia2021}), the relative orbital motion for wide binary systems can be robustly measured by Gaia across a range of distances and total system masses, even assuming that the \textit{relative} proper motion uncertainties are at least a factor of $\sqrt{2}$ higher than the values above. 

In terms of quantities reported by Gaia, the relative proper motion of the binary system in au~yr$^{-1}$ is 
\begin{equation}
    \mu=\sqrt{\left[\Delta\left({\pi^{-1}\dot{\alpha}\cos\delta}\right)\right]^{2} + \left[\Delta\left(\pi^{-1}\dot{\delta}\right)\right]^2}
\label{eqn:mu}
\end{equation}
assuming known parallax $\pi$ and proper motions $\dot{\alpha}\cos{\delta}$ and $\dot{\delta}$ for both stars in the binary. This is the projection on the sky of the three-dimensional motion of one star relative to another. In general, this projected speed is a complex function of the eccentricity, phase, and orientation of the orbit to our line of sight (see Appendix A of \citealt{Pearce_2020} for a derivation of the projected relative orbital motion in terms of Keplerian orbital elements). For a circular, face-on orbit, $\mu$ will be constant, but in the general case $\mu$ is not a good estimate of the actual relative orbital velocity of the binary system. However, just as with $\gamma$, the statistical properties of $\mu$ for a population of systems can be used to make inferences about the overall properties of those systems, including total mass and eccentricity. 

\cite{Tokovinin2016} described a quantity $\mu'$ that is the ratio of $\mu$ to the expected orbital motion assuming that the system is circular and observed face-on, which we denote $\mu^*$ ($\mu'=\mu/\mu^{*}$). Following from Kepler's third law, $\mu^*=2\pi s^{-1/2}~M_\mathrm{tot}^{1/2}$, where $s$ is the current projected separation of the binary pair, in au, and $M_\mathrm{tot}$ is the total system mass, in solar masses, and the resulting units are au~yr$^{-1}$ assuming distance is known directly through a parallax measurement. A gravitationally bound binary star system with a Keplerian orbit will have $0\leq\mu'<\sqrt{2}$. For a given system, the distribution of $\mu'$ across all orbital phases can be simulated. Or, for a population of systems assumed to have the same $M_\mathrm{tot}$ and eccentricities drawn from the same distribution, the distribution we define as $D\equiv\mu' M_\mathrm{tot}^{1/2}$ (which is what is measured using astrometric data from Gaia), can be used to constrain mass and eccentricity for the population.

We simulate the distribution of $\mu'$, or $P(\mu'|\alpha)$ for known $M_\mathrm{tot}$, where $\alpha$ is a parameter that describes the overall distribution of orbital eccentricities as 
\begin{equation}
P\left(e|\alpha\right)=\left(1+\alpha\right)e^\alpha.
\label{eq:e_given_alpha}
\end{equation}

\begin{figure}[h]
    \centering
    \includegraphics[width=\linewidth]{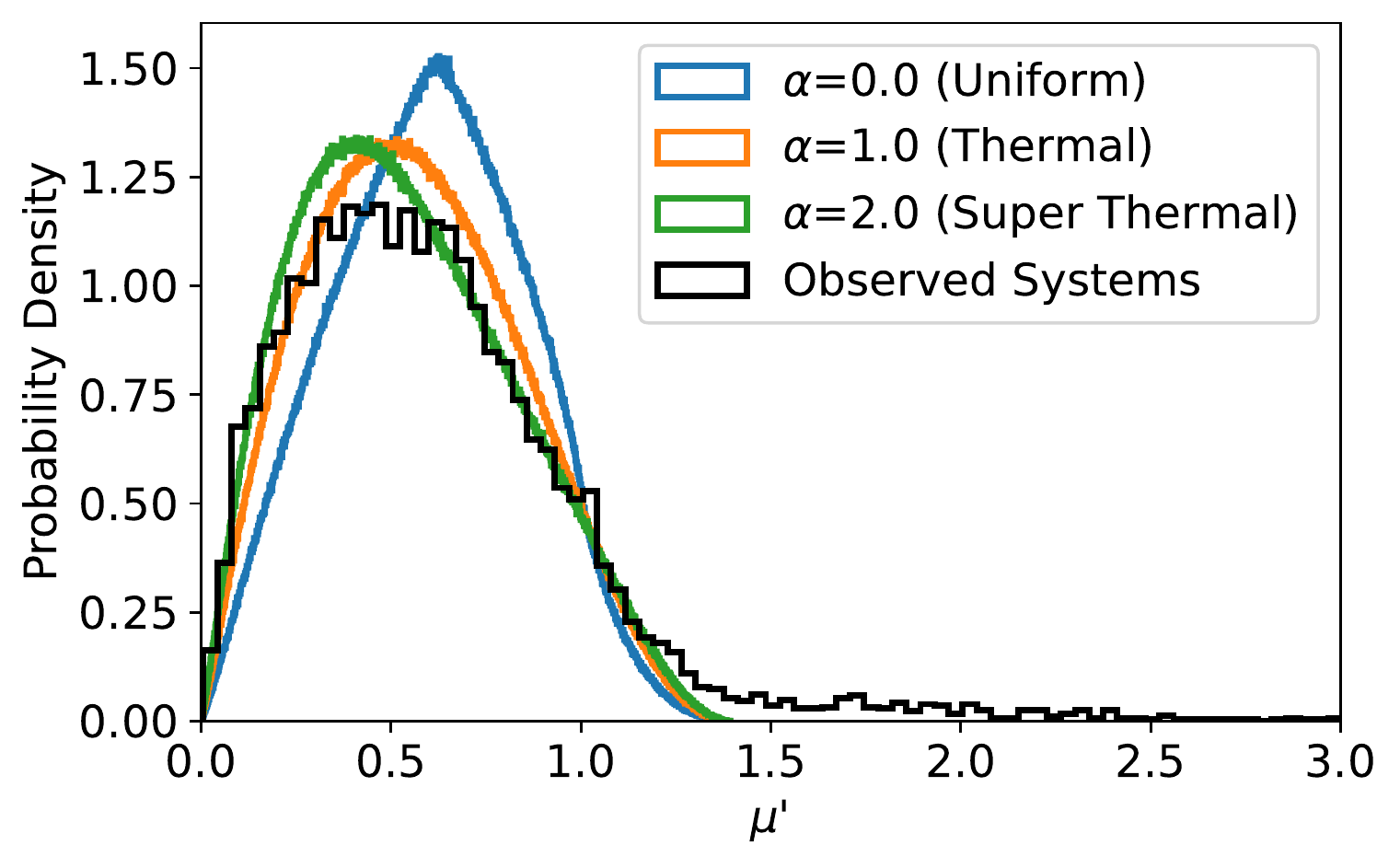}
    \caption{$P(\mu'|\alpha)$ for three different values of $\alpha$. The distribution for our observed population of binary star systems in the cases where $M_\mathrm{tot}$ can be estimated from the \citet{Mann_2019} relation is given in black. The tail beyond $\mu'>\sqrt{2}$, which is non-physical for bound binary systems, is thought to be the result of intrinsic measurement error and hierarchical star systems misinterpreted as binaries.}
    \label{fig:muprime_alpha}
\end{figure}

We simulate $\mu'$ for $10^6$ random Keplerian systems with orbital eccentricities drawn from $P(e|\alpha)$, inclinations drawn from a uniform distribution in $\cos{i}$, argument of periastron, $\omega$, drawn from a uniform distribution between 0 and $\pi$, and orbital phase drawn uniformly between 0 and $2\pi$. Following Appendix A of \citet{Pearce_2020}, we find that $\mu'$ does not explicitly depend on the longitude of the ascending node, $\Omega$. We assume a log-normal period distribution from \citet{Raghavan2010}, though we note that $P(\mu'|\alpha)$ does not actually depend on the distribution of orbital periods. \cite{Hwang2022} used the relationship between eccentricity and the relative motion angle $\gamma$ to investigate the eccentricity distribution for wide binary systems observed by Gaia. They found that the best-fit eccentricity distribution varies from consistent with uniform ($\alpha=0$) at small binary separations to super-thermal ($\alpha>1.0$) at larger separations of $>10^3$~au. So, we consider a grid of $0.0<\alpha<2.0$ in steps of 0.1. Our simulation assumes that $M_\mathrm{tot}$ is known, so the simulated distribution of $\mu'$ also does not depend on the distribution of $M_\mathrm{tot}$. The distribution $P(\mu'|\alpha)$ is the distribution of observed $\mu'$ for a population of binary systems if $M_\mathrm{tot}$ and $\alpha$ are known \textit{a priori}. Some examples of $P(\mu'|\alpha)$ are shown in Figure \ref{fig:muprime_alpha}. Changing $\alpha$ changes the overall shape, as well as the median and mean of the distribution of $\mu'$. Comparisons of $\mu'$ distributions for varying eccentricity parameterizations have been used in previous analyses as a probe to study observed populations of binary star systems (e.g., \citealt{2019MNRAS.488.4740P, 2022arXiv220502846P}).

\begin{figure}[h]
    \centering
    \includegraphics[width=\linewidth]{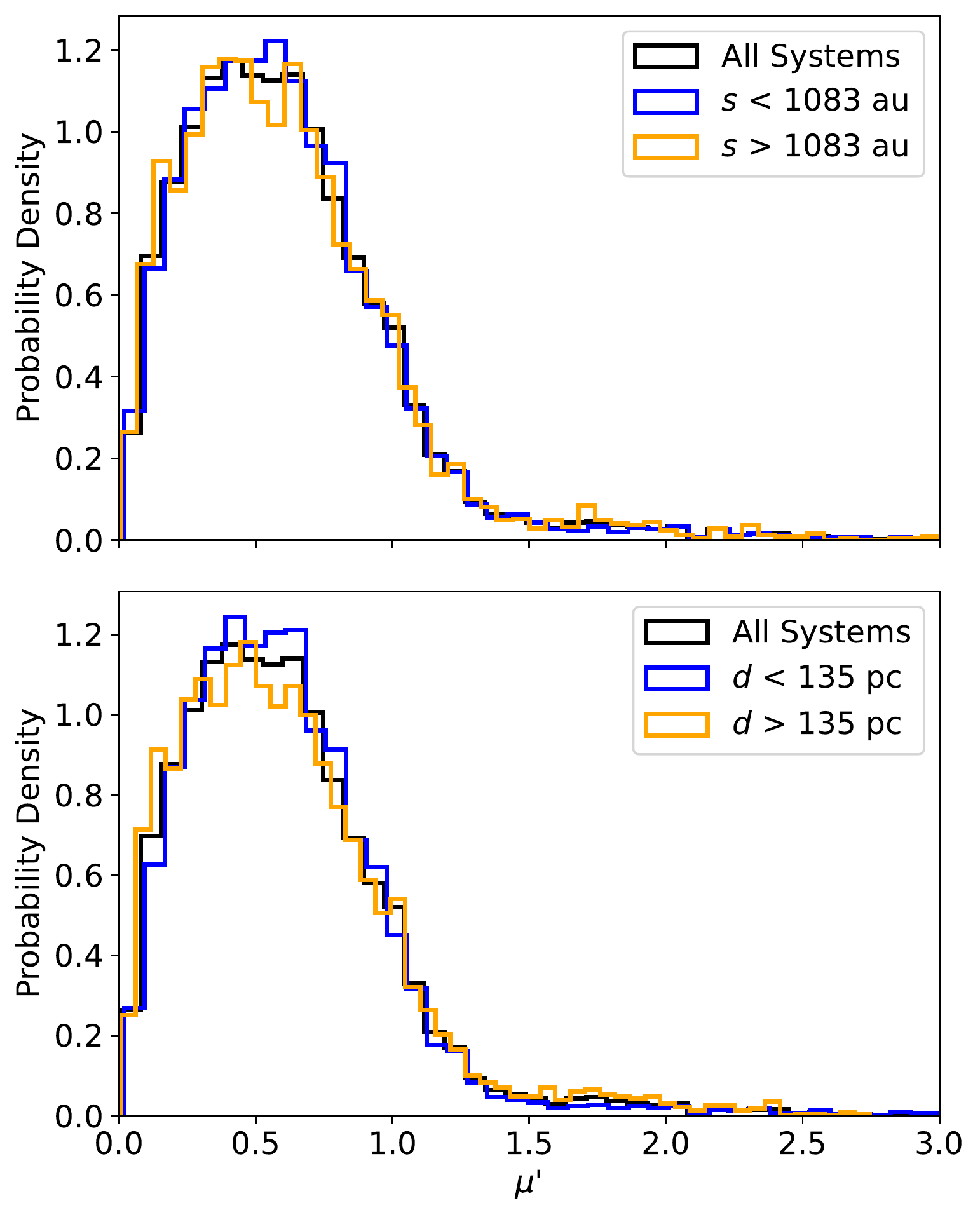}
    \caption{Comparing the observed $\mu'$ distributions for stars with \cite{Mann_2019} masses for systems with different physical separations and distances. Residual systematic effects in eDR3 proper motions may impact the relative proper motion measurement in systems where the components are closer together on the sky. We find broad agreement in the $\mu'$ distributions of these different subsamples. Top: $\mu'$ distribution for all systems, those with physical separations larger than 1083 au and smaller than 1083 au. Bottom: $\mu'$ distribution for all systems, those closer than 135 pc and farther than 135 pc.}
    \label{fig:muprime_alpha_twopanel}
\end{figure}

Despite the various cuts we make in selecting our sample, it is possible that undiagnosed systematic effects in eDR3 proper motions could bias our relative orbital motion measurements. Based on previous investigations of these issues in the literature \citep{cantat2021, Elbadry2021}, it is plausible that binary star systems that have tighter orbits or are closer to us may be more susceptible to these biases. We investigate possible separation-dependent systematic biases in our $\mu'$ distributions by comparing subsamples selected by physical separation and heliocentric distance. In Figure \ref{fig:muprime_alpha_twopanel}, we show the $\mu'$ distributions for our systems broken into bins of current physical separation (on either side of the median separation $s=1083$~au) and distance (on either side of the median distance $d=135$~pc) and find that the distributions of $\mu'$ are broadly consistent across these bins. Comparing the means of these distributions using a t-test, we find that the means (after excluding values $\mu'>1.5$) of the $s<1083$~au and $s>1083$~au $\mu'$ distributions, as well as the $d<135$~pc and $d>135$~pc $\mu'$ distributions, are consistent (null hypothesis of identical means has $p>0.05$ in both cases). We note that relative orbital motion biases in high-contrast systems may also be a concern, but these systems are excluded from our total system mass inference using our ``twin" sampling further described in Section \ref{sec:samp_select}.

By comparing the observed distribution of $\mu'$, which is derived from $D$ as a function of $M_\mathrm{tot}$, to $P(\mu'|\alpha)$ for a population of binary systems that are assumed to have the same $M_\mathrm{tot}$, we aim to directly estimate $M_\mathrm{tot}$ while marginalizing over $\alpha$. We will apply this technique to observations of binary star systems from Gaia to estimate an empirical Mass-Magnitude relation valid across the lower main sequence from $0.08~\mathrm{M_\odot}<M<1.0~\mathrm{M_\odot}$.

\section{Gaia Binaries - Sample Selection} \label{sec:samp_select}
With high-precision astrometry and photometry for nearly two billion stars, Gaia offers an unprecedented opportunity to study multiple star systems. \cite{Elbadry2021} carried out an analysis of the Gaia eDR3 database to identify pairs of stars having a high probability of being gravitationally bound based on common proper motion and proximity in three dimensions to produce a catalog of more than $10^6$ spatially resolved binaries. \cite{Elbadry2021} pare down the Gaia data set by considering only the systems that have parallaxes $>1~\mathrm{mas}$, fractional parallax uncertainties $<20\%$, absolute parallax uncertainties $<2~\mathrm{mas}$, and those where both stars have valid Gaia apparent $G$-band magnitudes. They also perform a robust analysis to estimate the likelihood that each system is a chance alignment (denoted as $\mathcal{R}$) rather than a true gravitationally bound system. 

From the \cite{Elbadry2021} catalog, we keep systems that satisfy the recommended threshold for having ``high bound probability" $\left(\mathcal{R} < 0.1\right)$, therefore removing pairs that have a non-negligible likelihood of being chance alignments. We follow \cite{Fabricius2021} and \cite{daltio2021} by removing systems where either star has re-normalized unit-weight error \texttt{ruwe}~$>1.4$, \texttt{ipd\_frac\_multi\_peak}~$>10$, \texttt{ipd\_gof\_harmonic\_amplitude}~$>0.1$, or \texttt{astrometric\_excess\_noise~$>1.0$} all of which are known to be indicative of poor astrometric solutions. We exclude regions of high galactic extinction by requiring that $|b|>10^{\circ}$. We exclude white dwarfs by requiring that both stars in a pair have absolute magnitude $3<M_G<3.1\left(G_\mathrm{BP}-G_\mathrm{RP}\right)+5$. We also remove objects with apparent magnitude $G<5$ to avoid bright stars that may cause saturation. Since stars evolve with time, a Mass-Magnitude relation is generally more reliable for stars that evolve very slowly. For example, in the Gaia H-R diagrams shown in \citet{gaiahr2018} and \citet{daltio2021}, stars with absolute $G$ magnitude $M_G=2.0$ could be massive main sequence stars or lower mass stars that have evolved onto the giant branch. However, the lowest mass main sequence stars evolve on timescales comparable to the age of the universe. For this reason, we select binary systems where both stars have absolute magnitudes $M_G>4.0$, corresponding approximately to $M<1.0~\mathrm{M}_{\sun}$.

We exclude binaries with angular separations $s~<~4~\arcsec$ from our analysis to mitigate issues with astrometric or photometric contamination. While this is larger than the formal Gaia confusion limit for systems with small magnitude differences, in binaries with larger magnitude differences the measurements of the fainter companion could be biased. \citet{brandeker2019} simulated source detection effects and found that Gaia should have good sensitivity to systems with contrasts as large as $\Delta G=8.0$ at separations of $4\arcsec$. We apply a cut on the photometric measurement error by requiring that both stars in a pair have \texttt{phot\_rp\_mean\_flux\_over\_error~$>10$}. The reason we only enforce this photometric signal-to-noise cut in the $G_\mathrm{RP}$ passband is because we choose Gaia's RP filter to build our mass-luminosity relationship from. We chose to develop our Mass-Magnitude relation in the red Gaia $G_\mathrm{RP}$ band, instead of the broad $G$ band, for three reasons. First, the lower main sequence stars we will investigate are relatively cool and have spectral energy distributions that peak at redder wavelengths ($G_{\rm{BP}}-G_{\rm{RP}}>1.0$). Second, the interstellar medium is more transparent to photons with redder wavelengths, so the $G_\mathrm{RP}$ photometry is less impacted by the effects of galactic extinction. Finally, we expect that the impact of metallicity on stellar flux is less pronounced at redder wavelengths (\citealt{MIST2016}). \citet{cantat2021} investigate magnitude-dependent biases in proper motions due to frame rotation in eDR3 and find a significant effect at $G=13$ that could bias our estimated relative orbital motion if the apparent magnitudes of the components straddle this boundary. To mitigate this effect, we exclude from our analyses any system where one component has $G<13.0$ while the other has $G>13.0$. Since we ultimately will analyze twin systems where both components have similar magnitudes, this cut only removes a small number of systems.

Following all of the astrometric and photometric cuts described in this section, we have a catalog 12,096 wide binary systems comprising lower main sequence stars $\left(M\lesssim1.0~\mathrm{M}_{\sun}\right)$ where both stars have high-quality Gaia measurements. For completeness, we re-simulate the $P(\mu'|\alpha)$ grid assigning random distances drawn from the actual distance distribution of these remaining Gaia wide binary systems and applying both the angular separation cut at $s>4\arcsec$ and a physical separation cut at $<5\times10^4$~au to the simulated systems. 

As we will discuss in Section \ref{sec:LLH}, to simplify the statistical framework for inferring the total system masses of the wide binaries in our sample, in some of the subsequent analyses we will further restrict our sample to systems where the stars have similar absolute magnitudes, with $\Delta M_{G_\mathrm{RP}} < 0.5$. These systems will also be restricted to a binning scheme in which only systems where both stars existing within a predefined set of arbitrary half-magnitude boundaries are kept. This cut further reduces our sample to 3,846 twin systems that will be used for the total mass inference, while also helping to reduce the impact of possible separation- or magnitude-dependent systematics of the type discussed above.

\section{Gaia Binaries - Measurements}\label{sec:measure}
In the general case where the total system mass is not known, we directly measure a quantity from the Gaia position, parallax, and proper motion measurements that we denote $D$, defined as
\begin{equation}
    D\equiv\mu' M_\mathrm{tot}^{1/2}=\frac{\mu}{2\pi} s^{1/2}.
\label{eqn:D}
\end{equation}
For each system, we estimate the uncertainty on $D$ through a Monte Carlo simulation, drawing $10^4$ samples from the reported Gaia measurements and uncertainties for right ascension, declination, the corresponding proper motion terms ($\dot{\alpha}\cos{\delta}$~and~$\dot{\delta}$), and the variance-weighted average parallax of the two stars to create an individual $D$ posterior for each system according to Equations \ref{eqn:mu} and \ref{eqn:D}. A small number of our systems have large angular separations, up to 30$'$, in which case projection effects in the observed angular motions of a physically co-moving pair of stars could be important. Following \citet{butkevish2014} we estimate the size of this effect given the actual locations and distances to the systems in our sample, and find that the differences in the observed proper motions were small, $<~1~\mathrm{\mu as}$~yr$^{-1}$ in almost all cases. Since the formal errors reported in eDR3 may be underestimated, we apply inflation factors to both the proper motions and parallaxes before calculating the statistical uncertainty on $D$ for each system. Following Figure 15 (right column) of \citet{Elbadry2021}, we inflate the reported parallax uncertainties by a factor of up to $\sim$1.3, depending on magnitude. For objects with apparent magnitudes $G<11$ we assume an inflation factor of 1.28, linearly decreasing to no error inflation at $G=18.0$. Following \citet{Brandt2021}, we apply an inflation factor of 1.37 to all the eDR3 proper motion uncertainties, independent of magnitude.

The uncertainties in $D$ for an individual system may be large given the magnitude of $\mu$ and the uncertainties in the Gaia proper motion and parallax measurements. For the purpose of selecting systems where $D$ is well-measured, we apply a cut on $\bar{\mu'}$, the expectation value of $\mu'$, and $\sigma_{\mu'}$, the uncertainty on $\mu'$, calculated from $D$ initially assuming $M_\mathrm{tot}=1~\mathrm{M_\odot}$. We exclude systems from our sample that have $\bar{\mu'}>3$, which is non-physical for bound binary systems, and $\sigma_{\mu'}>0.1$. This is effectively a selection on distance and apparent magnitude, which primarily determine the proper motion and parallax errors. Since $\sigma_{\mu'}$ only depends on Gaia astrometric uncertainties, and not on system orbital parameters, this cut does not bias the Mass-Magnitude relation we wish to derive. After making the cuts described in Sections \ref{sec:samp_select} and \ref{sec:measure}, we are left with a sample of 12,096 binary star systems, of which 3,846 have $\Delta M_{G_\mathrm{RP}}<0.5$, are within the bins described in Section \ref{sec:samp_select}, and are therefore in our twin subsample.

\section{Mass Inference} \label{sec:LLH}
An initial investigation of the measured values of $\mu'$ for the subset of our Gaia sample that is bright enough to have 2MASS $K_s$ magnitudes, and therefore potential mass estimates through the \citet{Mann_2019} Mass-Magnitude relation, revealed a significant population of systems (at least 10$\%$) with $\mu'$ significantly larger than the $\mu'=\sqrt{2}$ upper limit for bound, Keplerian orbits as seen in the simulations described in Section \ref{sec:muprime} and shown in Figure~\ref{fig:muprime_alpha}. It is possible that these systems are not actually bound, though given the relative proximity of these systems to the Sun and the detailed analyses carried out in \citet{Elbadry2021}, this seems unlikely to be a significant source of high $\mu'$ values. Alternatively, these systems could be higher-order multiples built from one or more unresolved binaries. Similarly, Jupiter-mass companions with semi-major axes of a few au orbiting either or both stars in a wide binary could induce apparent relative proper motion of a similar magnitude to the Keplerian motion of the wide binary itself. In these cases, the photometric mass estimate will in general be underestimated since, given the slope of the mass-luminosity relation, a relative increase in mass results in a much smaller relative increase in luminosity. \citet{Raghavan2010} explored multiplicity in a nearly volume-limited sample of stars and estimated that up to 25\% of apparent wide binary systems may in fact be triple or quadruple systems (see Table 16 in \citealt{Raghavan2010}). From long-term Doppler exoplanet surveys, we know that Jupiter-mass companions with semi-major axes of a few au are somewhat rare, occurring around approximately 6\% of Sun-like stars in the survey presented by \citet{wittenmyer2020}. While a tail of systems with $\mu'>1.4$ is therefore expected (cases where mass is \textit{underestimated} based on absolute magnitude), it remains  difficult to explain systems with $\mu'\sim3.0$, which in this scenario would require the \citet{Mann_2019} photometric mass to be low by a factor of at least four. Independent of the physical explanation for this tail, it is clear that we will need to include this in our model of the observed $\mu'$ distribution.

By comparing the observed $\mu'$ distribution for the cases where both stars in a binary have mass estimates from the \citet{Mann_2019} relation, we found that the overall distribution is well-described by the sum of our simulated $P(\mu'|\alpha)$ and an additional half Gaussian centered at zero. Following the results of \citet{Hwang2022} for wide binary systems ($s>1000$~au for almost all of our systems), we began by assuming that $\alpha=1.3$. This introduces two new free parameters into our model: $N$, the overall normalization of the tail population and $\sigma$, the width of the additional Gaussian. The overall model is then
\begin{eqnarray}
    \label{eq:model}
    \begin{aligned}
    P(\mu'|\alpha, N, \sigma)    = (1-N) P(\mu'|\alpha)+
    N\frac{\sqrt{2}}{\sigma\sqrt{\pi}}e^{-\frac{\mu'^{2}}{2\sigma^2}}.
    \end{aligned}
    \label{eqn:model}
\end{eqnarray}
As discussed in Section \ref{sec:samp_select}, we focus our analyses on absolute $G_\mathrm{RP}$ magnitudes. To simplify our analysis, we sort the binary catalog to identify twin systems, which we define to have similar magnitudes ($\Delta G_\mathrm{RP}~<~0.5$~mag) and fall within the confines of our magnitude bins described in Section \ref{sec:samp_select}. Given the slope of the \citet{Mann_2019} $K_s$-band Mass-Magnitude relation, the variation in mass over a bin is expected to be small, less than $0.1$~M$_{\sun}$ per 0.5 magnitudes at 0.5~M$_{\sun}$. Our analysis could also be carried out by inferring the parameters of an empirical function that maps absolute $G_{\rm{RP}}$ magnitude to mass, therefore increasing the total number of systems in our sample by alleviating the half-magnitude binning constraint. However, in this approach the parameters of the Mass-Magnitude relation are effectively latent variables. This, combined with the fact that the masses of the two stars in a binary system may be correlated (that is to say the system mass ratio, $q$, is not necessarily uniformly distributed; see \citealt{elbadryq} and Section \ref{sec:apps} below), significantly complicates the statistical inference framework. Given the large total number of systems in our sample, the twin approach provides a straightforward path to estimating a Mass-Magnitude relation.


Following Bayes' Theorem, we wish to infer the posterior probability of the model given the data (we directly measure $D$ with Gaia, so that is the data), which in this case is
\begin{equation}
    \label{eq:Bayes}
    \resizebox{.905\hsize}{!}{$P(M_\mathrm{tot},\alpha, N, \sigma | D) \propto P(D|M_\mathrm{tot},\alpha, N, \sigma) P(M_\mathrm{tot},\alpha, N, \sigma)$.}
\end{equation}
Here, the second term on the right side of the proportionality contains our priors on the free parameters of the model. The first term on the right side of the proportionality is a likelihood that can be calculated directly from our measurements of $D$, as well as our simulations of $\mu'$ described in Section \ref{sec:muprime} and the model in Equation \ref{eq:model}. Given that for a fixed set of values for $\alpha$, $N$, and $\sigma$, $P(D|M_{\mathrm{tot}})=P(\mu'| M_\mathrm{tot}^{1/2})$, the likelihood can be expressed as the sum of the log-likelihoods of the individual $D_i$ measurements within an $M_{G_{\rm{RP}}}$ bin as
\begin{equation}
\begin{aligned}
\mathcal{L} = \sum_{i}\ln \left[P(D_{i}|M_\mathrm{tot}^{1/2}, \alpha, N, \sigma) \right].
\end{aligned}
\end{equation}
We divide our binary twin sample into 16 equally spaced bins across the range $12.0>M_{G_\mathrm{RP}}>4.0$ and carry out a Markov Chain Monte Carlo analysis to sample from the posterior probabilities on $M_\mathrm{tot}$, $\alpha$, $N$, and $\sigma$ given the systems in each bin. Here, we are implicitly assuming that within each bin $M_\mathrm{tot}$ is twice the mass of either individual star. We employ a Metropolis-Hastings algorithm and generate chains with $10^5$ steps, removing the initial $10^3$ steps as burn-in. We linearly interpolate within the simulated grid of $P(\mu'|\alpha)$ described in Section \ref{sec:muprime} to evaluate $P(\mu'|\alpha)$ for specific values of $\alpha$ and enforce normalization of the model from Equation \ref{eq:model} for each set of trial parameters. We assume uniform priors on all free parameters as follows: $0.0<\alpha<2.0$, $0.0~\mathrm{M_\odot}<M_\mathrm{tot}<3.0~\mathrm{M_\odot}$, $0.0<N<0.5$, and $0.1<\sigma<3.0$. Examples of fits to individual $M_{G_\mathrm{RP}}$ bins are shown in Figure \ref{fig:P_D}. For each bin, we calculate the modal value of the $M_\mathrm{tot}$ posterior along with a 68\% highest density credible interval. The resulting individual mass estimates are hereafter denoted as $M_\mathrm{GORP}~=~M_\mathrm{tot}/2$.

For the lowest two mass bins we consider, $11.5~>~M_{G_\mathrm{RP}}~>~11.0$ and $12.0~>~M_{G_\mathrm{RP}}~>~11.5$, there are too few twin systems in our wide binary sample for our analysis to work, so instead we consider cases where one star has absolute magnitude within one of these low-luminosity bins and the other has $10.5>M_{G_\mathrm{RP}}>10.0$, fixing the mass of the more massive star as determined above. For these systems, we include the statistical uncertainty derived from the posterior on stellar mass for the more massive stars in the estimation of the credible interval for the mass of the less massive star. 

We evaluate the robustness of our modeling framework to the influence of unseen companions, either stellar or sub-stellar, through a simulation that injects random additional relative motion into our $\mu$ measurements to assess the impact on the resulting estimates of $M_\mathrm{tot}$. As discussed earlier in this section, an unseen companion will tend to  bias our mass estimates by increasing the measured relative orbital motion. We find that for additional stellar motion at the level of up to 0.04 au yr$^{-1}$, our mass estimates are not significantly biased at the level of our statistical uncertainties.
 
\section{Results} \label{sec:results}
\begin{figure*}[t]
    \centering
    \includegraphics[width=\textwidth]{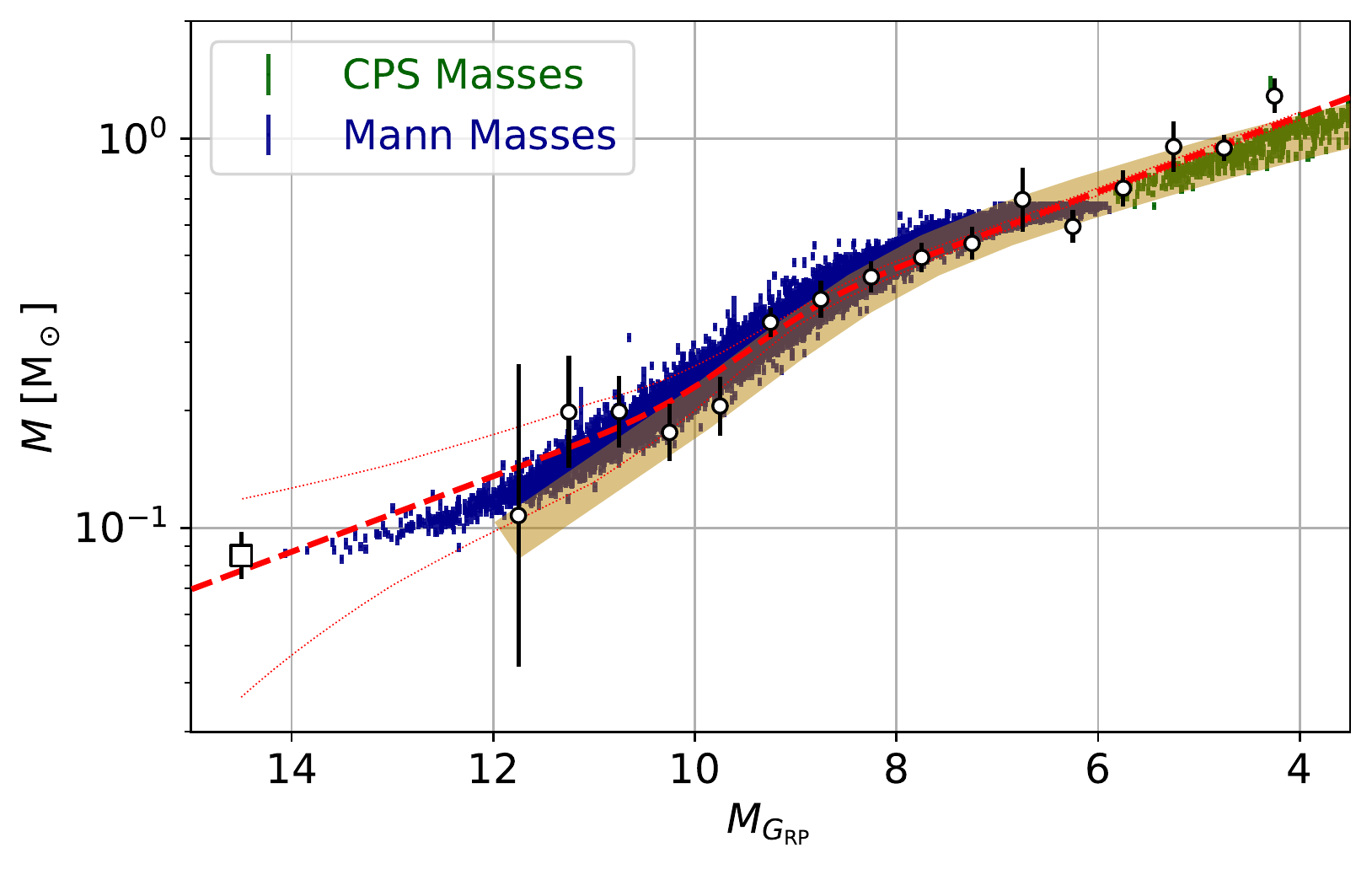}
    \caption{Stellar mass as a function of absolute $G_\mathrm{RP}$ magnitude. Stars from our sample of binary systems that have 2MASS measurements with $10.5~>~M_{K_s}~>~4.5$ and $K_{s}<14.0$ can have their masses estimated via the \cite{Mann_2019} relation and are shown in blue. Green points show the masses derived from isochrones and high-resolution spectra by \cite{brewer2016}. Our best-fit $M_\mathrm{GORP}$ mass estimates and corresponding uncertainties are shown for each bin in absolute $G_\mathrm{RP}$ magnitude as white points at each bin center. The best-fit Mass-Magnitude relation from Equation \ref{eqn:modlinear} is shown as the red dashed line and the 68$\%$ confidence interval for that relation is shown by the red dotted lines. The MIST evolutionary track for 5~Gyr solar metallicity stars is overplotted in gold for reference. The square point at $M_{G_\mathrm{RP}}=14.5$ is from the literature and is not derived from our analyses, and as such is denoted with a square.}
    \label{fig:mass_mag}
\end{figure*}

\begin{figure}[h]
    \centering
    \includegraphics[width=\linewidth]{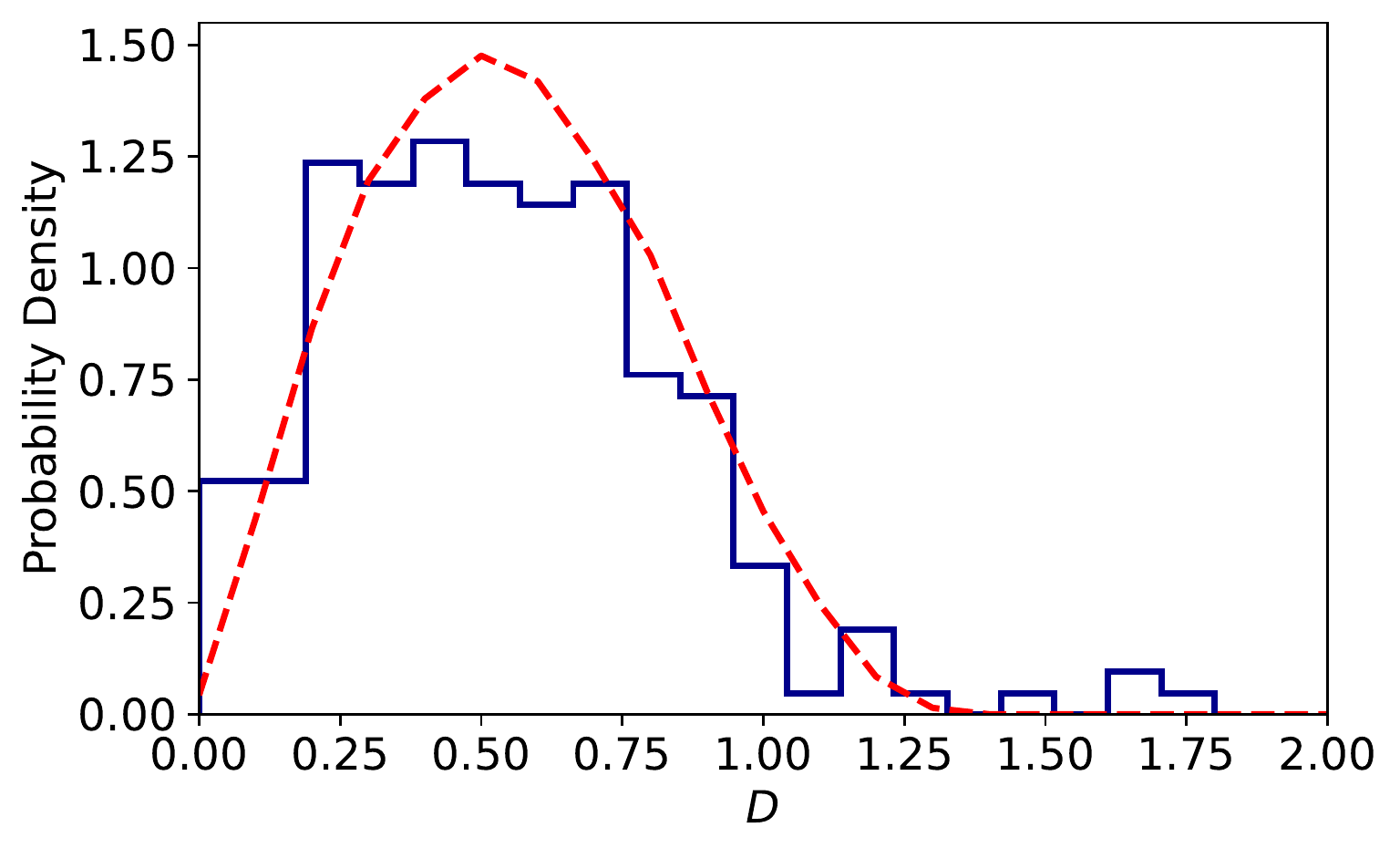}
    \caption{Distribution of $D$ within the $8.5~>~M_{G_\mathrm{RP}}~>~8.0$ bin. The best-fit model for these data is indicated by the dashed line. For this particular set of 223 twin binary systems, the best-fit parameters for the model given in Equation \ref{eqn:model} are $M_\mathrm{tot}=0.88\pm0.08~\mathrm{M}_\odot$, $\alpha=0.90\pm0.49$, $N=0.22\pm0.09$, and $\sigma=0.68\pm0.28$.}
    \label{fig:P_D}
\end{figure}

\begin{figure}[h]
    \centering
    \includegraphics[width=\linewidth]{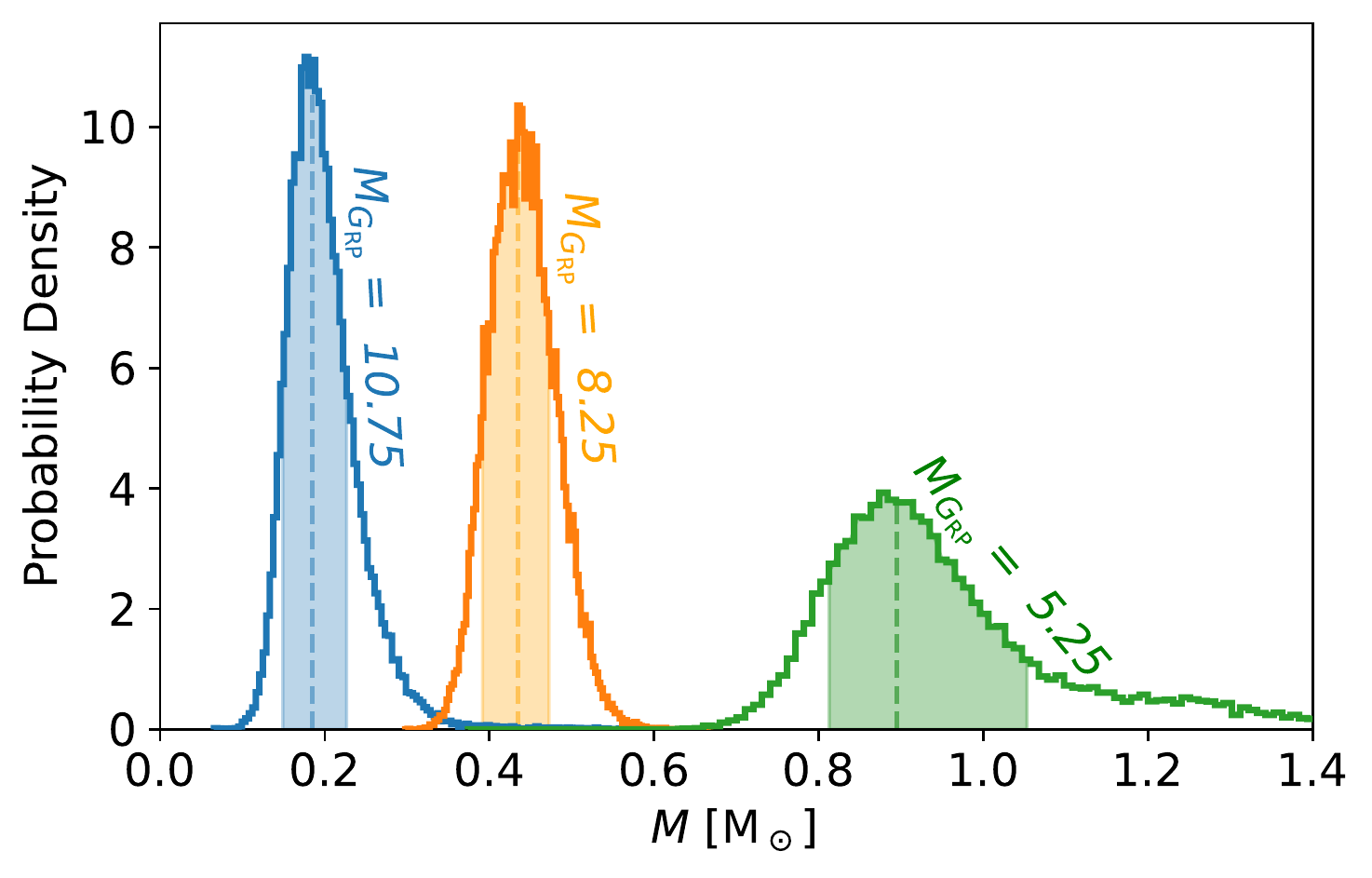}
    \caption{Sample posterior distributions for three of our specified absolute magnitude bins. The 68\% highest density interval is shown underneath the distribution with the modal value indicated with a dashed line.}
    \label{fig:MGRP_posteriors}
\end{figure}

In Figure \ref{fig:mass_mag} we compare our individual estimates of $M_\mathrm{GORP}$ based on the twin sample across 16 bins in $M_{G_{\rm{RP}}}$ to mass estimates based on the \citet{Mann_2019} $K_s$ Mass-Magnitude relation, MIST evolutionary models at 5~Gyr from \citet{MIST2016}, and 1,617 isochrone-based mass estimates from \citet{brewer2016}. We find excellent agreement with our $M_\mathrm{GORP}$ mass estimates given the typical statistical mass uncertainties derived from the posteriors on $M_\mathrm{GORP}$ of $\sim10\%$ for $M>0.02M~\mathrm{M}_\odot$, rising to $>25\%$ at lower masses. Examples of individual posterior distributions and their 68\% highest credible density intervals for three of our absolute magnitude bins are shown in Figure \ref{fig:MGRP_posteriors}.

For the parameter $N$, the normalization of the tail component of our model, we find consistent values of $N=0.22\pm0.07$ across the mass range we consider. While the physical interpretation of this is difficult, we note that this value is consistent with the relative frequencies of double and triple systems found in \citet{Raghavan2010}. Our data do not place strong constraints on $\alpha$, but across our mass range we find that $\alpha$ is consistent with $1.19\pm0.56$. While \cite{Hwang2022} found in their analyses that $\alpha$ was constrained to approximately $1.25\pm0.25$ for wide ($>1000$~au) systems, we note that their analysis involved a much broader range of binary types and was not broken down by stellar spectral type, mass, or evolutionary state. Given this, our results on the eccentricity distribution appear broadly consistent with \cite{Hwang2022}. For the parameter $\sigma$, the width of the tail distribution, we find $\sigma=0.85\pm0.30$ with no correlation with $M_\mathrm{GORP}$.

We evaluated analytic relations between absolute $G_\mathrm{RP}$ magnitude and the $M_\mathrm{GORP}$ masses. Because of the small number of faint binary systems where both stars pass our photometric and astrometric cuts, our sample only includes systems with $M_{G_\mathrm{RP}}<12.0$, but hydrogen-burning stars can be significantly fainter. For the purposes of these fits, we follow \citet{rayle2018} and fix the magnitude of an object near the hydrogen burning limit with a mass of $0.08\pm0.01$~M$_{\sun}$ at absolute magnitude of $M_{G_\mathrm{RP}}=14.5$. There is a relatively sharp break in our estimated masses around $M_{G_\mathrm{RP}}=9.5$ that makes it difficult to fit our results with a single low-order polynomial. Empirically, we found that our data are well-fitted by the following modified linear function, which enforces a smooth transition to lower masses at $M_{G_\mathrm{RP}}>9.5$ by using the error function, erf(x). 
\begin{equation}
    \resizebox{.905\hsize}{!}{$
    \log_{10}(M_\mathrm{GORP}) = a + b M_{G_\mathrm{RP}}-c \left[1+\mathrm{erf}\left(M_{G_\mathrm{RP}}-9.5\right)\right]$}
\label{eqn:modlinear}
\end{equation}
While this functional form is not physically motivated, it describes the data well with fewer degrees of freedom than a comparable polynomial relation. Based on a $\Delta \rm{AIC}>1.1$, this modified linear model is slightly preferred over a polynomial fit of first order alone (and strongly preferred over higher-order polynomial fits - see \citealt{ivezic2019}). The posteriors on $\log_{10}\left(M_\mathrm{GORP}\right)$ are approximately Gaussian, so we use least-squares optimization to estimate the best-fit values for the three model parameters in Equation~\ref{eqn:model}. We use a Markov Chain Monte Carlo approach to generate 10$^{5}$ samples from the posteriors of the three parameters using a Metropolis-Hastings sampler. Based on the modal values of the posteriors we find $a=0.445$, $b=-0.097$, and $c=0.075$ valid over the range $14.5>M_{G_\mathrm{RP}}>4.0$. Given the covariances between these parameters, the uncertainty in the Mass-Magnitude relation is not well-described by the uncertainties on the parameters as derived from the individual posteriors. Point estimates and corresponding uncertainties based on Monte Carlo simulations are given in Table \ref{table:1}. These values can be interpolated to provide point estimates for mass within the range $14.5>M_{G_\mathrm{RP}}>4.0$. 

\begin{deluxetable}{lcccccccccccc}
\tablecaption{Mass-Magnitude Relation}
\label{table:1}
\startdata
\tablehead{\colhead{$M_{G_\mathrm{RP}}$} &&&&&& \colhead{$\log_{10}M$ (M$_{\sun}$)} &&&&&& \colhead{$\sigma_{\log_{10}M}$}}\\[-0.4cm]
4.0 &&&&&& 0.058 &&&&&& 0.028 \\
5.0 &&&&&& -0.039 &&&&&& 0.019 \\
6.0 &&&&&& -0.136 &&&&&& 0.015\\
7.0 &&&&&& -0.233 &&&&&& 0.015 \\
8.0 &&&&&& -0.333 &&&&&& 0.020 \\
9.0 &&&&&& -0.463 &&&&&& 0.016 \\
10.0 &&&&&& -0.637 &&&&&& 0.030\\
11.0 &&&&&& -0.778 &&&&&& 0.040 \\
12.0 &&&&&& -0.867 &&&&&& 0.037 \\
13.0 &&&&&& -0.964 &&&&&& 0.036 \\
14.5 &&&&&& -1.110 &&&&&& 0.039 \\
\enddata
\tablecomments{Point estimates and uncertainties for $\log_{10}{M}$ based on the modified linear Mass-Magnitude relation described in Section \ref{sec:results}.}
\end{deluxetable}


In \citet{Mann_2019} and other examples of Mass-Magnitude relations in the literature (e.g. \citealt{1993AJ....106..773H, Delfosse2000}), photometric mass estimates are typically calibrated against a small set of stars (62 in the case of \citealt{Mann_2019}) that have precise dynamical mass estimates. We searched the literature for visual binary systems that are wide enough to be well-resolved by Gaia and have precise (better than $5\%$ precision) mass measurements, either total for the system or for individual components. Unfortunately, we identified fewer than a dozen such systems. 

\begin{figure}[h]
    \centering
    \includegraphics[width=\linewidth]{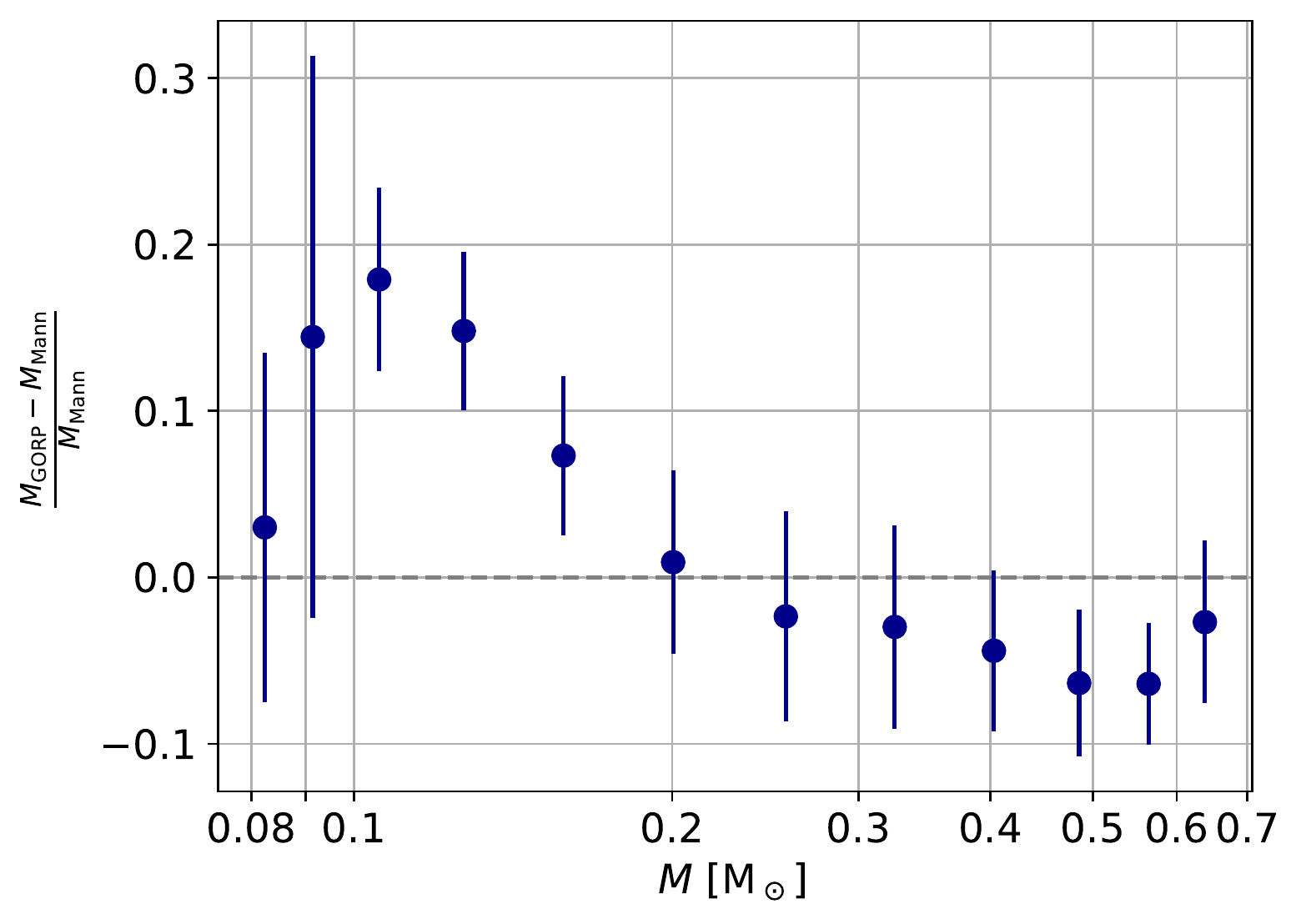}
    \caption{Relative errors in our $M_\mathrm{GORP}$ mass estimates compared to those calculated from the \cite{Mann_2019} method. The relative internal uncertainties, $\sigma_\mathrm{GORP}/M_\mathrm{Mann}$, are shown as error bars.}
    \label{fig:gorp_mann_rel_err}
\end{figure}

The strong agreement between $M_\mathrm{GORP}$ and $M_\mathrm{Mann}$ is clear from Figure \ref{fig:mass_mag}. Since the \citet{Mann_2019} masses are expected to have very small internal errors, at the level of $\sim2\%$, by directly comparing $M_\mathrm{GORP}$ and $M_\mathrm{Mann}$ we can assess the internal errors of the $M_\mathrm{GORP}$ estimates derived here. Using the same photometric and astrometric quality cuts described in Section \ref{sec:samp_select}, we identified a sample of individual Gaia stars that also had 2MASS photometry in the absolute magnitude range recommended by \cite{Mann_2019} for making reliable mass estimates $\left(10.5>M_{K_{s}}>4.5\right)$. Additionally, we only selected stars with $K_s<14.0$ to mitigate the effects of photometric errors on the resulting mass estimates. In total, we selected $\sim$30,000 stars and calculated the distribution of the quantity $\Delta M \equiv M_\mathrm{GORP}-M_\mathrm{Mann}$ across 12 absolute $K_{s}$ magnitude bins spanning the aforementioned range, each 0.5 magnitude in width. The distribution of $\Delta M$ within each bin is approximately Gaussian, so its standard deviation is equivalent to the sum in quadrature of $\sigma_\mathrm{GORP}$ and $\sigma_\mathrm{Mann}$. Therefore, we estimate the internal uncertainty in our relation as $\sigma_\mathrm{GORP}=\left[\sigma_{\Delta M}^2 - \sigma_\mathrm{Mann}^2\right]^{1/2}$. As shown in Figure \ref{fig:gorp_mann_rel_err}, we find that the uncertainty in our mass estimates relative to \citet{Mann_2019}, $\sigma_\mathrm{GORP}/M_\mathrm{Mann}$, is approximately $5-10\%$ over most of our mass range, though it increases below 0.1~M$_{\sun}$. These values are consistent with the mass posteriors derived in Section \ref{sec:LLH}. For $M\lesssim 0.1~\mathrm{M}_\odot$, our $M_\mathrm{GORP}$ mass estimates are somewhat biased high, at the $\sim15\%$ level, relative to \citet{Mann_2019}. 

\section{Applications}\label{sec:apps}
The substantially greater depth of Gaia compared to 2MASS allows us to extend an accurate Mass-Magnitude relation for low-mass stars over a much larger volume than the \citet{Mann_2019} relation alone, particularly for the lowest mass stars. There are many potential applications of this relation, including estimating the field mass function and studying the relationship between stellar mass and orbital parameters in binary systems. 

\begin{figure}[h]
    \centering
    \includegraphics[width=\linewidth]{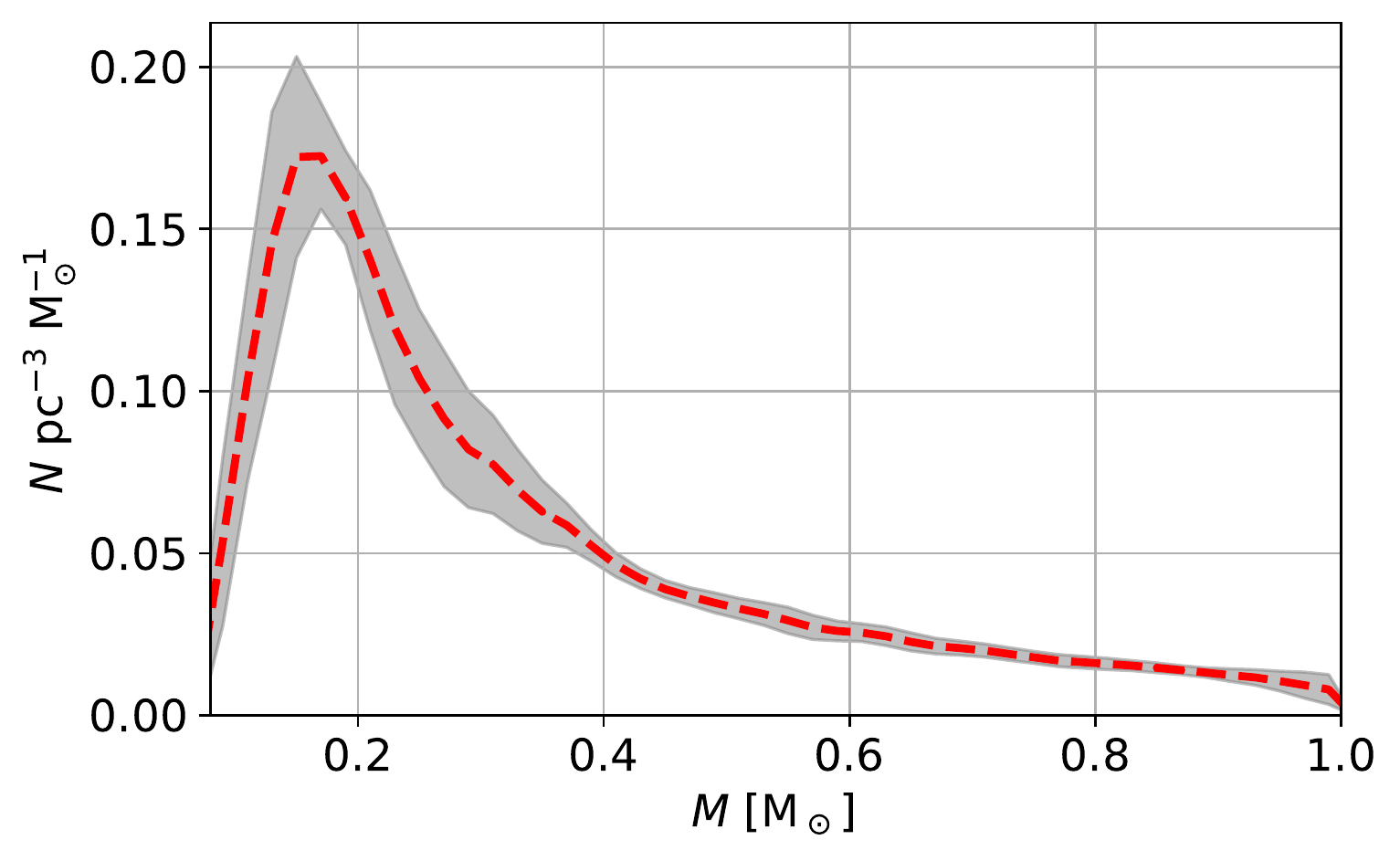}
    \caption{Mass function estimated using our $M_\mathrm{GORP}$ relation over the volume-limited Gaia sample of stars in the range of $10~\mathrm{pc}<d<50~\mathrm{pc}$ and $|b|>10^{\circ}$. The mean number density of stars per cubic parsec per solar mass is shown in red, along with the 68$\%$ confidence interval.} 
    \label{fig:mass_function}
\end{figure}

We apply the same astrometric and photometric cuts described in Section \ref{sec:samp_select} to select a volume-limited sample of (apparently) single lower main sequence stars with $10~\mathrm{pc}<d<50~\mathrm{pc}$. The inner limit was chosen to avoid issues related to saturation for the brightest stars in our sample, and the outer limit was chosen so that the faintest main sequence stars, down to $M_{G_\mathrm{RP}}=14.5$, will be relatively bright and detected with \texttt{phot\_rp\_mean\_flux\_over\_error~$>10$}, and the parallaxes for all of the objects will be well-measured. We use our derived Mass-Magnitude relation (Equation \ref{eqn:modlinear}) and associated parameter uncertainties to estimate the field mass function between $0.1~\mathrm{M}_{\sun}<M<1.0~\mathrm{M}_{\sun}$ in units of Stars pc$^{-3}$ M$_{\sun}^{-1}$ as shown in Figure \ref{fig:mass_function}. In this calculation, we ignore the photometric and parallax uncertainties for individual stars, which for the subsample considered here are small compared to the uncertainty in the Mass-Magnitude relation. We find that the field mass function peaks at approximately 0.16~M$_{\sun}$, broadly consistent with other results in the literature including \citet{kroupa93}, \citet{rana1987}, and \citet{sollima2019}. 

We have not made an attempt to correct this mass function for the effects of unresolved binaries. Following the results of \citet{Raghavan2010} by assuming an overall multiple fraction of 40\% in the stellar mass range considered here, integrating over a log-normal distribution of period (see Figure 13 of \citealt{Raghavan2010}), and assuming a typical total system mass of $M_\mathrm{tot}=1.0$~M$_\odot$, approximately 25\% of our stars could be unresolved binaries given the Gaia small-separation detection limits described in \citet{zeigler2018} and \citet{brandeker2019}. The effect of unresolved binaries on our mass function is difficult to quantify. While there is some evidence that wide binaries may be preferentially of equal mass ( \citealt{elbadryq}), the unresolved systems in our sample will on average have much smaller semi-major axes than the resolved binary systems. At the same time, on the lower main sequence a companion near or below the hydrogen-burning limit will contribute negligibly to the total flux of the system. So, in our analysis an unresolved binary will have an overestimated mass, by up to 25\% for an equal-mass system given the slope of our Mass-Magnitude relation, and will result in counting one slightly more massive star instead of two lower mass stars. \citet{sollima2019} approached estimating the mass function of a similar sample of lower main sequence stars using isochrones and a stellar multiplicity model to forward-model the observed Gaia color-magnitude diagram. The results presented here are consistent with the much more sophisticated analysis of \citet{sollima2019} in the limit of an unresolved binary fraction less than $40\%$, which is what we expect based on previous binarity surveys (see Figure 4 in \citealt{sollima2019}). As noted by \citet{sollima2019}, it is also possible that some unresolved binaries with photocentric motion or complex point spread functions are excluded from our sample by the various astrometric quality cuts we apply. 

Our Mass-Magnitude relation can also be used to study the properties of the wide binary systems in our sample. For example, the distribution of binary mass ratios could be used to constrain theories and simulations of star formation. While some authors have found that the binary mass ratio, $q$, is approximately uniformly distributed for certain populations of binary systems (for example, \citealt{Raghavan2010} or \citealt{tokovinin2014}), others have found evidence for a distribution that peaks at smaller values of $q$ (see \citealt{DM1991}, \citealt{gullikson2016}, \citealt{moe2017}), or a distribution that has a clear excess of systems at $q=1.0$ (see \citealt{elbadryq}). In each of these cases, different populations of systems, covering different total system masses and separations and with different selection biases, are being studied. 

\begin{figure}[h]
    \centering    \includegraphics[width=\linewidth]{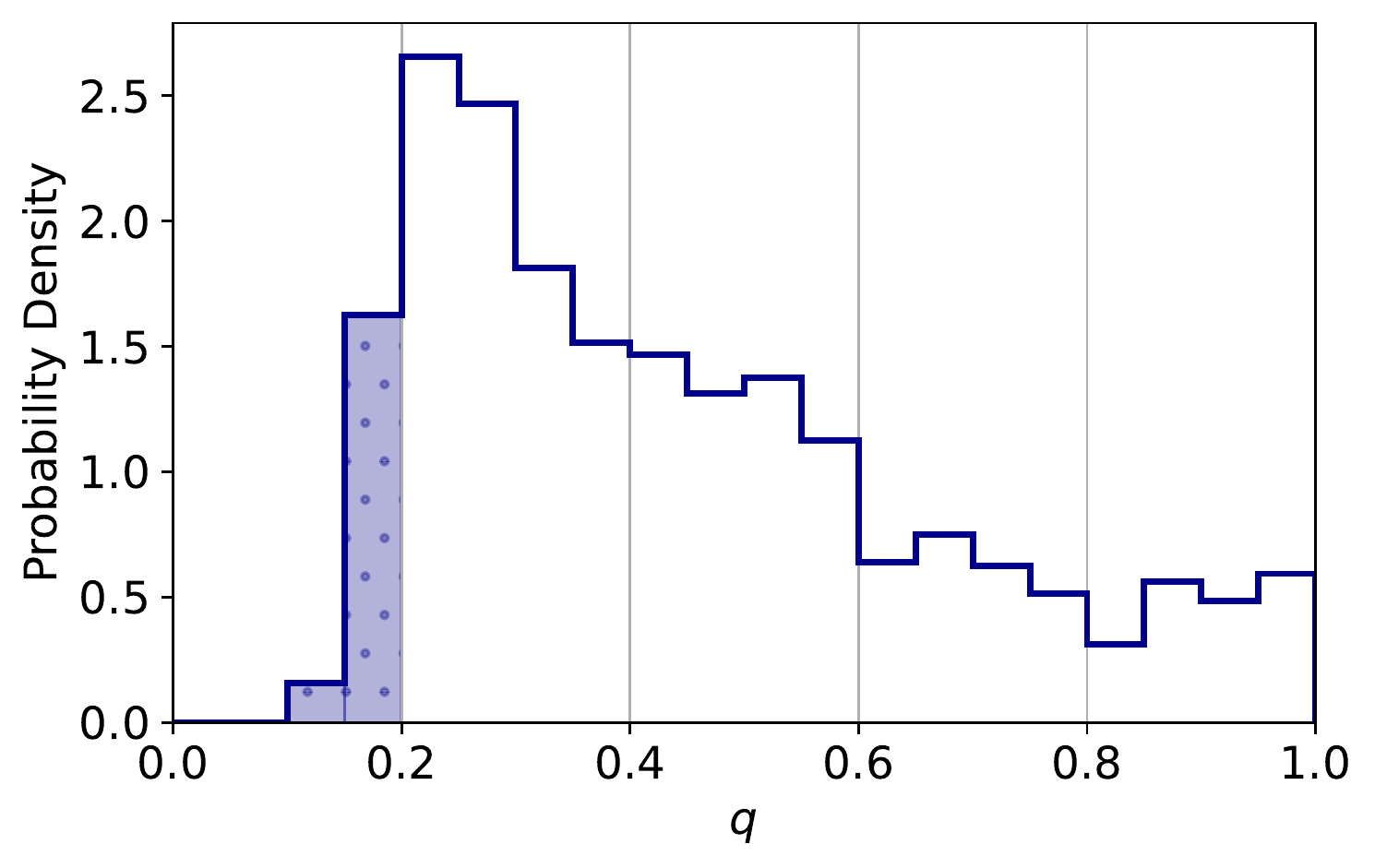}
    \caption{Distribution of mass ratios $q=M_2/M_1$ inferred from our Mass-Magnitude relation using binary star systems with $0.8~\mathrm{M}_\odot<M<0.9~\mathrm{M}_\odot$ within $d<210$~pc. We find that the distribution of $q$ increases toward lower values of $q$. Our sample is not necessarily volume-limited for $q<0.2$, denoted by the shaded region.} 
    \label{fig:q_sep}
\end{figure}

In an effort to study mass ratios in a sample of wide binaries that is relatively free from selection biases, we select binaries from our catalog that have projected separations $>100$~au, are within 210~pc, and have a primary (more massive) star in the mass range $0.8~\mathrm{M}_\odot<M<0.9~\mathrm{M}_\odot$ (early-K spectral type) using all of the photometric and astrometric quality cuts described in Section \ref{sec:samp_select}, but without the cut on $\sigma_{\mu'}<0.1$. Following \citet{brandeker2019}, we further select systems with angular separations $s>8\arcsec$ to avoid detection biases in high-contrast systems. By selecting distance less than 210~pc and primary stars in this mass range, we ensure that, given our photometric quality cuts, we are sensitive to all systems with mass ratios $q>0.2$. Based on 1,506 wide binary systems passing our selection criteria, we calculate the distribution of $q$ as shown in Figure \ref{fig:q_sep}. We find that the distribution of mass ratios peaks at $q=0.2$ and decreases towards $q=1.0$, which for K-type primaries is consistent with \cite{2013MNRAS.430L...6G}, though we note that our sample is not necessarily volume-limited for mass ratios below this. Since our sample is approximately volume-limited down to $M_{G_\mathrm{RP}}=10.75$, we do not expect the angular separation cut at $s>8\arcsec$ to produce different average total (or primary) stellar masses across our range in projected separation.

\section{Conclusion} \label{sec:concl}

Precise estimates of stellar masses are key to our understanding of a range of scientific topics from the physical characteristics of exoplanets to the formation and evolution of stars. However, these measurements remain difficult to make even today and typically rely on time-consuming observations of short-period binary star benchmark systems. On the lower main sequence, where stars evolve relatively slowly, it is possible to infer mass from luminosity alone. We derive a Mass-Magnitude relation using the Gaia $G_\mathrm{RP}$ passband. Compared to previously published Mass-Magnitude relations based on 2MASS or $V$-band photometry of relatively bright stars and tied to a small number of precise dynamical measurements, our approach is internally calibrated using the orbital motions of a large number of wide binary systems and is applicable over the large volume probed by Gaia.

Starting from the catalog of wide binary star systems derived from Gaia eDR3 by \citet{Elbadry2021}, we apply a number of cuts to ensure high-quality photometric and astrometric measurements and select a sample of 12,096 binary systems across the lower main sequence and analyze the subset of 3,846 pairs having similar magnitudes ($\Delta M_{G_\mathrm{RP}}<0.5)$. For these systems, we measure the projected orbital motion of one star relative to the other as the difference in the measured proper motions, and compare this to simulations of projected orbital motion to infer the total mass of the twin systems (assumed to be twice the mass of the individual stars) in bins of absolute $G_\mathrm{RP}$ magnitude while marginalizing over the eccentricity distribution of the population. Using a Bayesian framework, we estimate stellar mass as a function of absolute $G_\mathrm{RP}$ magnitude for $12.0>M_{G_\mathrm{RP}}>4.0$ with typical internal uncertainties of $5-25\%$ depending on the mass. Assuming a literature value for the absolute magnitude of a star at the hydrogen-burning limit, we fit for a modified linear relation between $M_{G_\mathrm{RP}}$ and $\log_{10}({M_\mathrm{GORP}})$ across the range $14.5>M_{G_\mathrm{RP}}>4.0$.

Our $M_{G_\mathrm{RP}}$ range encompasses the absolute magnitude range of the Mass-Magnitude relation presented by \cite{Mann_2019}, which allows us to directly compare our $M_\mathrm{GORP}$ masses to $M_\mathrm{Mann}$ masses with a sample of stars having measured $K_s$ magnitudes from 2MASS within the magnitude range recommended by \citet{Mann_2019}. To do this, we cross-matched Gaia and 2MASS to find all well-measured stars with $10.5>M_{K_s}>4.5$. We find that our mass estimates for individual objects are consistent with those from \cite{Mann_2019} at the $5-10\%$ level for $M>0.1~\mathrm{M}_{\sun}$, though we note that our masses appear to be somewhat biased relative to \citet{Mann_2019} in the $M<0.2~\mathrm{M}_{\sun}$ range. The Mass-Magnitude relation presented here extends the reach of photometric mass estimation to more than 30 million individual stars in the Gaia eDR3 database, nearly an order of magnitude more objects than in the sample where the \citet{Mann_2019} relation can be directly applied. We use our results to estimate the field stellar mass function in the solar neighborhood for stars within $10-50$~pc of the Sun down to the hydrogen-burning limit, and find that the mass function peaks at a mass of $0.16~\mathrm{M}_{\sun}$, consistent with previous results in the literature. We also use our Mass-Magnitude relation to study the mass ratios of wide binary systems with early K-type dwarf primaries. For this population, we find that the distribution of $q$ is not consistent with uniform, rather decreasing toward $q=1.0$.

\begin{acknowledgments}
We thank the referee, Tim Brandt, for insightful comments that helped to improve this manuscript. We also thank Gary Bernstein, Mike Jarvis, Bhuvnesh Jain, Cyrille Doux, and Marco Raveri for helpful conversations that improved this work. MRG would like to acknowledge support from the NSF through a Graduate Research Fellowship. 
\end{acknowledgments}

%

\vspace{5mm}
\facilities{Gaia, 2MASS}


\software{\texttt{astropy} \citep{astropy2013, astropy2018}}




\bibliography{sample631}{}

\begin{thebibliography}{}
\expandafter\ifx\csname natexlab\endcsname\relax\def\natexlab#1{#1}\fi
\providecommand{\url}[1]{\href{#1}{#1}}
\providecommand{\dodoi}[1]{doi:~\href{http://doi.org/#1}{\nolinkurl{#1}}}
\providecommand{\doeprint}[1]{\href{http://ascl.net/#1}{\nolinkurl{http://ascl.net/#1}}}
\providecommand{\doarXiv}[1]{\href{https://arxiv.org/abs/#1}{\nolinkurl{https://arxiv.org/abs/#1}}}

\bibitem[{{Astropy Collaboration} {et~al.}(2013){Astropy Collaboration},
  {Robitaille}, {Tollerud}, {Greenfield}, {Droettboom}, {Bray}, {Aldcroft},
  {Davis}, {Ginsburg}, {Price-Whelan}, {Kerzendorf}, {Conley}, {Crighton},
  {Barbary}, {Muna}, {Ferguson}, {Grollier}, {Parikh}, {Nair}, {Unther},
  {Deil}, {Woillez}, {Conseil}, {Kramer}, {Turner}, {Singer}, {Fox}, {Weaver},
  {Zabalza}, {Edwards}, {Azalee Bostroem}, {Burke}, {Casey}, {Crawford},
  {Dencheva}, {Ely}, {Jenness}, {Labrie}, {Lim}, {Pierfederici}, {Pontzen},
  {Ptak}, {Refsdal}, {Servillat}, \& {Streicher}}]{astropy2013}
{Astropy Collaboration}, {Robitaille}, T.~P., {Tollerud}, E.~J., {et~al.} 2013,
  \aap, 558, A33, \dodoi{10.1051/0004-6361/201322068}

\bibitem[{{Astropy Collaboration} {et~al.}(2018){Astropy Collaboration},
  {Price-Whelan}, {Sip{\H{o}}cz}, {G{\"u}nther}, {Lim}, {Crawford}, {Conseil},
  {Shupe}, {Craig}, {Dencheva}, {Ginsburg}, {VanderPlas}, {Bradley},
  {P{\'e}rez-Su{\'a}rez}, {de Val-Borro}, {Aldcroft}, {Cruz}, {Robitaille},
  {Tollerud}, {Ardelean}, {Babej}, {Bach}, {Bachetti}, {Bakanov}, {Bamford},
  {Barentsen}, {Barmby}, {Baumbach}, {Berry}, {Biscani}, {Boquien}, {Bostroem},
  {Bouma}, {Brammer}, {Bray}, {Breytenbach}, {Buddelmeijer}, {Burke},
  {Calderone}, {Cano Rodr{\'\i}guez}, {Cara}, {Cardoso}, {Cheedella}, {Copin},
  {Corrales}, {Crichton}, {D'Avella}, {Deil}, {Depagne}, {Dietrich}, {Donath},
  {Droettboom}, {Earl}, {Erben}, {Fabbro}, {Ferreira}, {Finethy}, {Fox},
  {Garrison}, {Gibbons}, {Goldstein}, {Gommers}, {Greco}, {Greenfield},
  {Groener}, {Grollier}, {Hagen}, {Hirst}, {Homeier}, {Horton}, {Hosseinzadeh},
  {Hu}, {Hunkeler}, {Ivezi{\'c}}, {Jain}, {Jenness}, {Kanarek}, {Kendrew},
  {Kern}, {Kerzendorf}, {Khvalko}, {King}, {Kirkby}, {Kulkarni}, {Kumar},
  {Lee}, {Lenz}, {Littlefair}, {Ma}, {Macleod}, {Mastropietro}, {McCully},
  {Montagnac}, {Morris}, {Mueller}, {Mumford}, {Muna}, {Murphy}, {Nelson},
  {Nguyen}, {Ninan}, {N{\"o}the}, {Ogaz}, {Oh}, {Parejko}, {Parley}, {Pascual},
  {Patil}, {Patil}, {Plunkett}, {Prochaska}, {Rastogi}, {Reddy Janga},
  {Sabater}, {Sakurikar}, {Seifert}, {Sherbert}, {Sherwood-Taylor}, {Shih},
  {Sick}, {Silbiger}, {Singanamalla}, {Singer}, {Sladen}, {Sooley},
  {Sornarajah}, {Streicher}, {Teuben}, {Thomas}, {Tremblay}, {Turner},
  {Terr{\'o}n}, {van Kerkwijk}, {de la Vega}, {Watkins}, {Weaver}, {Whitmore},
  {Woillez}, {Zabalza}, \& {Astropy Contributors}}]{astropy2018}
{Astropy Collaboration}, {Price-Whelan}, A.~M., {Sip{\H{o}}cz}, B.~M., {et~al.}
  2018, \aj, 156, 123, \dodoi{10.3847/1538-3881/aabc4f}

\bibitem[{{Brandeker} \& {Cataldi}(2019)}]{brandeker2019}
{Brandeker}, A., \& {Cataldi}, G. 2019, \aap, 621, A86,
  \dodoi{10.1051/0004-6361/201834321}

\bibitem[{{Brandt}(2021)}]{Brandt2021}
{Brandt}, T.~D. 2021, \apjs, 254, 42, \dodoi{10.3847/1538-4365/abf93c}

\bibitem[{Brandt {et~al.}(2019)Brandt, Dupuy, \& Bowler}]{Brandt2019}
Brandt, T.~D., Dupuy, T.~J., \& Bowler, B.~P. 2019, The Astronomical Journal,
  158, 140, \dodoi{10.3847/1538-3881/ab04a8}

\bibitem[{{Brewer} {et~al.}(2016){Brewer}, {Fischer}, {Valenti}, \&
  {Piskunov}}]{brewer2016}
{Brewer}, J.~M., {Fischer}, D.~A., {Valenti}, J.~A., \& {Piskunov}, N. 2016,
  \apjs, 225, 32, \dodoi{10.3847/0067-0049/225/2/32}

\bibitem[{{Butkevich} \& {Lindegren}(2014)}]{butkevish2014}
{Butkevich}, A.~G., \& {Lindegren}, L. 2014, \aap, 570, A62,
  \dodoi{10.1051/0004-6361/201424483}

\bibitem[{{Cantat-Gaudin} \& {Brandt}(2021)}]{cantat2021}
{Cantat-Gaudin}, T., \& {Brandt}, T.~D. 2021, \aap, 649, A124,
  \dodoi{10.1051/0004-6361/202140807}

\bibitem[{{Choi} {et~al.}(2016){Choi}, {Dotter}, {Conroy}, {Cantiello},
  {Paxton}, \& {Johnson}}]{MIST2016}
{Choi}, J., {Dotter}, A., {Conroy}, C., {et~al.} 2016, \apj, 823, 102,
  \dodoi{10.3847/0004-637X/823/2/102}

\bibitem[{{Dal Tio} {et~al.}(2021){Dal Tio}, {Mazzi}, {Girardi}, {Barbieri},
  {Zaggia}, {Bressan}, {Chen}, {Costa}, \& {Marigo}}]{daltio2021}
{Dal Tio}, P., {Mazzi}, A., {Girardi}, L., {et~al.} 2021, \mnras, 506, 5681,
  \dodoi{10.1093/mnras/stab1964}

\bibitem[{{Delfosse} {et~al.}(2000){Delfosse}, {Forveille}, {S{\'e}gransan},
  {Beuzit}, {Udry}, {Perrier}, \& {Mayor}}]{Delfosse2000}
{Delfosse}, X., {Forveille}, T., {S{\'e}gransan}, D., {et~al.} 2000, \aap, 364,
  217.
\newblock \doarXiv{astro-ph/0010586}

\bibitem[{{Duquennoy} \& {Mayor}(1991)}]{DM1991}
{Duquennoy}, A., \& {Mayor}, M. 1991, \aap, 248, 485

\bibitem[{{El-Badry} {et~al.}(2021){El-Badry}, {Rix}, \&
  {Heintz}}]{Elbadry2021}
{El-Badry}, K., {Rix}, H.-W., \& {Heintz}, T.~M. 2021, \mnras, 506, 2269,
  \dodoi{10.1093/mnras/stab323}

\bibitem[{{El-Badry} {et~al.}(2019){El-Badry}, {Rix}, {Tian}, {Duch{\^e}ne}, \&
  {Moe}}]{elbadryq}
{El-Badry}, K., {Rix}, H.-W., {Tian}, H., {Duch{\^e}ne}, G., \& {Moe}, M. 2019,
  \mnras, 489, 5822, \dodoi{10.1093/mnras/stz2480}

\bibitem[{{Fabricius} {et~al.}(2021){Fabricius}, {Luri}, {Arenou}, {Babusiaux},
  {Helmi}, {Muraveva}, {Reyl{\'e}}, {Spoto}, {Vallenari}, {Antoja}, {Balbinot},
  {Barache}, {Bauchet}, {Bragaglia}, {Busonero}, {Cantat-Gaudin}, {Carrasco},
  {Diakit{\'e}}, {Fabrizio}, {Figueras}, {Garcia-Gutierrez}, {Garofalo},
  {Jordi}, {Kervella}, {Khanna}, {Leclerc}, {Licata}, {Lambert}, {Marrese},
  {Masip}, {Ramos}, {Robichon}, {Robin}, {Romero-G{\'o}mez}, {Rubele}, \&
  {Weiler}}]{Fabricius2021}
{Fabricius}, C., {Luri}, X., {Arenou}, F., {et~al.} 2021, \aap, 649, A5,
  \dodoi{10.1051/0004-6361/202039834}

\bibitem[{{Gaia Collaboration} {et~al.}(2016){Gaia Collaboration}, {Prusti},
  {de Bruijne}, {Brown}, {Vallenari}, {Babusiaux}, {Bailer-Jones}, {Bastian},
  {Biermann}, {Evans}, {Eyer}, {Jansen}, {Jordi}, {Klioner}, {Lammers},
  {Lindegren}, {Luri}, {Mignard}, {Milligan}, {Panem}, {Poinsignon},
  {Pourbaix}, {Randich}, {Sarri}, {Sartoretti}, {Siddiqui}, {Soubiran},
  {Valette}, {van Leeuwen}, {Walton}, {Aerts}, {Arenou}, {Cropper}, {Drimmel},
  {H{\o}g}, {Katz}, {Lattanzi}, {O'Mullane}, {Grebel}, {Holland}, {Huc},
  {Passot}, {Bramante}, {Cacciari}, {Casta{\~n}eda}, {Chaoul}, {Cheek}, {De
  Angeli}, {Fabricius}, {Guerra}, {Hern{\'a}ndez}, {Jean-Antoine-Piccolo},
  {Masana}, {Messineo}, {Mowlavi}, {Nienartowicz}, {Ord{\'o}{\~n}ez-Blanco},
  {Panuzzo}, {Portell}, {Richards}, {Riello}, {Seabroke}, {Tanga},
  {Th{\'e}venin}, {Torra}, {Els}, {Gracia-Abril}, {Comoretto},
  {Garcia-Reinaldos}, {Lock}, {Mercier}, {Altmann}, {Andrae}, {Astraatmadja},
  {Bellas-Velidis}, {Benson}, {Berthier}, {Blomme}, {Busso}, {Carry},
  {Cellino}, {Clementini}, {Cowell}, {Creevey}, {Cuypers}, {Davidson}, {De
  Ridder}, {de Torres}, {Delchambre}, {Dell'Oro}, {Ducourant}, {Fr{\'e}mat},
  {Garc{\'\i}a-Torres}, {Gosset}, {Halbwachs}, {Hambly}, {Harrison}, {Hauser},
  {Hestroffer}, {Hodgkin}, {Huckle}, {Hutton}, {Jasniewicz}, {Jordan},
  {Kontizas}, {Korn}, {Lanzafame}, {Manteiga}, {Moitinho}, {Muinonen},
  {Osinde}, {Pancino}, {Pauwels}, {Petit}, {Recio-Blanco}, {Robin}, {Sarro},
  {Siopis}, {Smith}, {Smith}, {Sozzetti}, {Thuillot}, {van Reeven}, {Viala},
  {Abbas}, {Abreu Aramburu}, {Accart}, {Aguado}, {Allan}, {Allasia},
  {Altavilla}, {{\'A}lvarez}, {Alves}, {Anderson}, {Andrei}, {Anglada Varela},
  {Antiche}, {Antoja}, {Ant{\'o}n}, {Arcay}, {Atzei}, {Ayache}, {Bach},
  {Baker}, {Balaguer-N{\'u}{\~n}ez}, {Barache}, {Barata}, {Barbier}, {Barblan},
  {Baroni}, {Barrado y Navascu{\'e}s}, {Barros}, {Barstow}, {Becciani},
  {Bellazzini}, {Bellei}, {Bello Garc{\'\i}a}, {Belokurov}, {Bendjoya},
  {Berihuete}, {Bianchi}, {Bienaym{\'e}}, {Billebaud}, {Blagorodnova},
  {Blanco-Cuaresma}, {Boch}, {Bombrun}, {Borrachero}, {Bouquillon}, {Bourda},
  {Bouy}, {Bragaglia}, {Breddels}, {Brouillet}, {Br{\"u}semeister},
  {Bucciarelli}, {Budnik}, {Burgess}, {Burgon}, {Burlacu}, {Busonero}, {Buzzi},
  {Caffau}, {Cambras}, {Campbell}, {Cancelliere}, {Cantat-Gaudin}, {Carlucci},
  {Carrasco}, {Castellani}, {Charlot}, {Charnas}, {Charvet}, {Chassat},
  {Chiavassa}, {Clotet}, {Cocozza}, {Collins}, {Collins}, {Costigan}, {Crifo},
  {Cross}, {Crosta}, {Crowley}, {Dafonte}, {Damerdji}, {Dapergolas}, {David},
  {David}, {De Cat}, {de Felice}, {de Laverny}, {De Luise}, {De March}, {de
  Martino}, {de Souza}, {Debosscher}, {del Pozo}, {Delbo}, {Delgado},
  {Delgado}, {di Marco}, {Di Matteo}, {Diakite}, {Distefano}, {Dolding}, {Dos
  Anjos}, {Drazinos}, {Dur{\'a}n}, {Dzigan}, {Ecale}, {Edvardsson}, {Enke},
  {Erdmann}, {Escolar}, {Espina}, {Evans}, {Eynard Bontemps}, {Fabre},
  {Fabrizio}, {Faigler}, {Falc{\~a}o}, {Farr{\`a}s Casas}, {Faye}, {Federici},
  {Fedorets}, {Fern{\'a}ndez-Hern{\'a}ndez}, {Fernique}, {Fienga}, {Figueras},
  {Filippi}, {Findeisen}, {Fonti}, {Fouesneau}, {Fraile}, {Fraser}, {Fuchs},
  {Furnell}, {Gai}, {Galleti}, {Galluccio}, {Garabato}, {Garc{\'\i}a-Sedano},
  {Gar{\'e}}, {Garofalo}, {Garralda}, {Gavras}, {Gerssen}, {Geyer}, {Gilmore},
  {Girona}, {Giuffrida}, {Gomes}, {Gonz{\'a}lez-Marcos},
  {Gonz{\'a}lez-N{\'u}{\~n}ez}, {Gonz{\'a}lez-Vidal}, {Granvik}, {Guerrier},
  {Guillout}, {Guiraud}, {G{\'u}rpide}, {Guti{\'e}rrez-S{\'a}nchez}, {Guy},
  {Haigron}, {Hatzidimitriou}, {Haywood}, {Heiter}, {Helmi}, {Hobbs},
  {Hofmann}, {Holl}, {Holland}, {Hunt}, {Hypki}, {Icardi}, {Irwin}, {Jevardat
  de Fombelle}, {Jofr{\'e}}, {Jonker}, {Jorissen}, {Julbe}, {Karampelas},
  {Kochoska}, {Kohley}, {Kolenberg}, {Kontizas}, {Koposov}, {Kordopatis},
  {Koubsky}, {Kowalczyk}, {Krone-Martins}, {Kudryashova}, {Kull}, {Bachchan},
  {Lacoste-Seris}, {Lanza}, {Lavigne}, {Le Poncin-Lafitte}, {Lebreton},
  {Lebzelter}, {Leccia}, {Leclerc}, {Lecoeur-Taibi}, {Lemaitre}, {Lenhardt},
  {Leroux}, {Liao}, {Licata}, {Lindstr{\o}m}, {Lister}, {Livanou}, {Lobel},
  {L{\"o}ffler}, {L{\'o}pez}, {Lopez-Lozano}, {Lorenz}, {Loureiro},
  {MacDonald}, {Magalh{\~a}es Fernandes}, {Managau}, {Mann}, {Mantelet},
  {Marchal}, {Marchant}, {Marconi}, {Marie}, {Marinoni}, {Marrese},
  {Marschalk{\'o}}, {Marshall}, {Mart{\'\i}n-Fleitas}, {Martino}, {Mary},
  {Matijevi{\v{c}}}, {Mazeh}, {McMillan}, {Messina}, {Mestre}, {Michalik},
  {Millar}, {Miranda}, {Molina}, {Molinaro}, {Molinaro}, {Moln{\'a}r},
  {Moniez}, {Montegriffo}, {Monteiro}, {Mor}, {Mora}, {Morbidelli}, {Morel},
  {Morgenthaler}, {Morley}, {Morris}, {Mulone}, {Muraveva}, {Musella},
  {Narbonne}, {Nelemans}, {Nicastro}, {Noval}, {Ord{\'e}novic},
  {Ordieres-Mer{\'e}}, {Osborne}, {Pagani}, {Pagano}, {Pailler}, {Palacin},
  {Palaversa}, {Parsons}, {Paulsen}, {Pecoraro}, {Pedrosa}, {Pentik{\"a}inen},
  {Pereira}, {Pichon}, {Piersimoni}, {Pineau}, {Plachy}, {Plum}, {Poujoulet},
  {Pr{\v{s}}a}, {Pulone}, {Ragaini}, {Rago}, {Rambaux}, {Ramos-Lerate},
  {Ranalli}, {Rauw}, {Read}, {Regibo}, {Renk}, {Reyl{\'e}}, {Ribeiro},
  {Rimoldini}, {Ripepi}, {Riva}, {Rixon}, {Roelens}, {Romero-G{\'o}mez},
  {Rowell}, {Royer}, {Rudolph}, {Ruiz-Dern}, {Sadowski}, {Sagrist{\`a}
  Sell{\'e}s}, {Sahlmann}, {Salgado}, {Salguero}, {Sarasso}, {Savietto},
  {Schnorhk}, {Schultheis}, {Sciacca}, {Segol}, {Segovia}, {Segransan},
  {Serpell}, {Shih}, {Smareglia}, {Smart}, {Smith}, {Solano}, {Solitro},
  {Sordo}, {Soria Nieto}, {Souchay}, {Spagna}, {Spoto}, {Stampa}, {Steele},
  {Steidelm{\"u}ller}, {Stephenson}, {Stoev}, {Suess}, {S{\"u}veges}, {Surdej},
  {Szabados}, {Szegedi-Elek}, {Tapiador}, {Taris}, {Tauran}, {Taylor},
  {Teixeira}, {Terrett}, {Tingley}, {Trager}, {Turon}, {Ulla}, {Utrilla},
  {Valentini}, {van Elteren}, {Van Hemelryck}, {van Leeuwen}, {Varadi},
  {Vecchiato}, {Veljanoski}, {Via}, {Vicente}, {Vogt}, {Voss}, {Votruba},
  {Voutsinas}, {Walmsley}, {Weiler}, {Weingrill}, {Werner}, {Wevers},
  {Whitehead}, {Wyrzykowski}, {Yoldas}, {{\v{Z}}erjal}, {Zucker}, {Zurbach},
  {Zwitter}, {Alecu}, {Allen}, {Allende Prieto}, {Amorim},
  {Anglada-Escud{\'e}}, {Arsenijevic}, {Azaz}, {Balm}, {Beck}, {Bernstein},
  {Bigot}, {Bijaoui}, {Blasco}, {Bonfigli}, {Bono}, {Boudreault}, {Bressan},
  {Brown}, {Brunet}, {Bunclark}, {Buonanno}, {Butkevich}, {Carret}, {Carrion},
  {Chemin}, {Ch{\'e}reau}, {Corcione}, {Darmigny}, {de Boer}, {de Teodoro}, {de
  Zeeuw}, {Delle Luche}, {Domingues}, {Dubath}, {Fodor}, {Fr{\'e}zouls},
  {Fries}, {Fustes}, {Fyfe}, {Gallardo}, {Gallegos}, {Gardiol}, {Gebran},
  {Gomboc}, {G{\'o}mez}, {Grux}, {Gueguen}, {Heyrovsky}, {Hoar}, {Iannicola},
  {Isasi Parache}, {Janotto}, {Joliet}, {Jonckheere}, {Keil}, {Kim},
  {Klagyivik}, {Klar}, {Knude}, {Kochukhov}, {Kolka}, {Kos}, {Kutka}, {Lainey},
  {LeBouquin}, {Liu}, {Loreggia}, {Makarov}, {Marseille}, {Martayan},
  {Martinez-Rubi}, {Massart}, {Meynadier}, {Mignot}, {Munari}, {Nguyen},
  {Nordlander}, {Ocvirk}, {O'Flaherty}, {Olias Sanz}, {Ortiz}, {Osorio},
  {Oszkiewicz}, {Ouzounis}, {Palmer}, {Park}, {Pasquato}, {Peltzer}, {Peralta},
  {P{\'e}turaud}, {Pieniluoma}, {Pigozzi}, {Poels}, {Prat}, {Prod'homme},
  {Raison}, {Rebordao}, {Risquez}, {Rocca-Volmerange}, {Rosen}, {Ruiz-Fuertes},
  {Russo}, {Sembay}, {Serraller Vizcaino}, {Short}, {Siebert}, {Silva},
  {Sinachopoulos}, {Slezak}, {Soffel}, {Sosnowska}, {Strai{\v{z}}ys}, {ter
  Linden}, {Terrell}, {Theil}, {Tiede}, {Troisi}, {Tsalmantza}, {Tur},
  {Vaccari}, {Vachier}, {Valles}, {Van Hamme}, {Veltz}, {Virtanen}, {Wallut},
  {Wichmann}, {Wilkinson}, {Ziaeepour}, \& {Zschocke}}]{Gaia2016}
{Gaia Collaboration}, {Prusti}, T., {de Bruijne}, J.~H.~J., {et~al.} 2016,
  \aap, 595, A1, \dodoi{10.1051/0004-6361/201629272}

\bibitem[{{Gaia Collaboration} {et~al.}(2018){Gaia Collaboration}, {Babusiaux},
  {van Leeuwen}, {Barstow}, {Jordi}, {Vallenari}, {Bossini}, {Bressan},
  {Cantat-Gaudin}, {van Leeuwen}, {Brown}, {Prusti}, {de Bruijne},
  {Bailer-Jones}, {Biermann}, {Evans}, {Eyer}, {Jansen}, {Klioner}, {Lammers},
  {Lindegren}, {Luri}, {Mignard}, {Panem}, {Pourbaix}, {Randich}, {Sartoretti},
  {Siddiqui}, {Soubiran}, {Walton}, {Arenou}, {Bastian}, {Cropper}, {Drimmel},
  {Katz}, {Lattanzi}, {Bakker}, {Cacciari}, {Casta{\~n}eda}, {Chaoul}, {Cheek},
  {De Angeli}, {Fabricius}, {Guerra}, {Holl}, {Masana}, {Messineo}, {Mowlavi},
  {Nienartowicz}, {Panuzzo}, {Portell}, {Riello}, {Seabroke}, {Tanga},
  {Th{\'e}venin}, {Gracia-Abril}, {Comoretto}, {Garcia-Reinaldos}, {Teyssier},
  {Altmann}, {Andrae}, {Audard}, {Bellas-Velidis}, {Benson}, {Berthier},
  {Blomme}, {Burgess}, {Busso}, {Carry}, {Cellino}, {Clementini}, {Clotet},
  {Creevey}, {Davidson}, {De Ridder}, {Delchambre}, {Dell'Oro}, {Ducourant},
  {Fern{\'a}ndez-Hern{\'a}ndez}, {Fouesneau}, {Fr{\'e}mat}, {Galluccio},
  {Garc{\'\i}a-Torres}, {Gonz{\'a}lez-N{\'u}{\~n}ez}, {Gonz{\'a}lez-Vidal},
  {Gosset}, {Guy}, {Halbwachs}, {Hambly}, {Harrison}, {Hern{\'a}ndez},
  {Hestroffer}, {Hodgkin}, {Hutton}, {Jasniewicz}, {Jean-Antoine-Piccolo},
  {Jordan}, {Korn}, {Krone-Martins}, {Lanzafame}, {Lebzelter}, {L{\"o}ffler},
  {Manteiga}, {Marrese}, {Mart{\'\i}n-Fleitas}, {Moitinho}, {Mora}, {Muinonen},
  {Osinde}, {Pancino}, {Pauwels}, {Petit}, {Recio-Blanco}, {Richards},
  {Rimoldini}, {Robin}, {Sarro}, {Siopis}, {Smith}, {Sozzetti}, {S{\"u}veges},
  {Torra}, {van Reeven}, {Abbas}, {Abreu Aramburu}, {Accart}, {Aerts},
  {Altavilla}, {{\'A}lvarez}, {Alvarez}, {Alves}, {Anderson}, {Andrei},
  {Anglada Varela}, {Antiche}, {Antoja}, {Arcay}, {Astraatmadja}, {Bach},
  {Baker}, {Balaguer-N{\'u}{\~n}ez}, {Balm}, {Barache}, {Barata}, {Barbato},
  {Barblan}, {Barklem}, {Barrado}, {Barros}, {Bartholom{\'e} Mu{\~n}oz},
  {Bassilana}, {Becciani}, {Bellazzini}, {Berihuete}, {Bertone}, {Bianchi},
  {Bienaym{\'e}}, {Blanco-Cuaresma}, {Boch}, {Boeche}, {Bombrun}, {Borrachero},
  {Bouquillon}, {Bourda}, {Bragaglia}, {Bramante}, {Breddels}, {Brouillet},
  {Br{\"u}semeister}, {Brugaletta}, {Bucciarelli}, {Burlacu}, {Busonero},
  {Butkevich}, {Buzzi}, {Caffau}, {Cancelliere}, {Cannizzaro}, {Carballo},
  {Carlucci}, {Carrasco}, {Casamiquela}, {Castellani}, {Castro-Ginard},
  {Charlot}, {Chemin}, {Chiavassa}, {Cocozza}, {Costigan}, {Cowell}, {Crifo},
  {Crosta}, {Crowley}, {Cuypers}, {Dafonte}, {Damerdji}, {Dapergolas}, {David},
  {David}, {de Laverny}, {De Luise}, {De March}, {de Martino}, {de Souza}, {de
  Torres}, {Debosscher}, {del Pozo}, {Delbo}, {Delgado}, {Delgado}, {Diakite},
  {Diener}, {Distefano}, {Dolding}, {Drazinos}, {Dur{\'a}n}, {Edvardsson},
  {Enke}, {Eriksson}, {Esquej}, {Eynard Bontemps}, {Fabre}, {Fabrizio},
  {Faigler}, {Falc{\~a}o}, {Farr{\`a}s Casas}, {Federici}, {Fedorets},
  {Fernique}, {Figueras}, {Filippi}, {Findeisen}, {Fonti}, {Fraile}, {Fraser},
  {Fr{\'e}zouls}, {Gai}, {Galleti}, {Garabato}, {Garc{\'\i}a-Sedano},
  {Garofalo}, {Garralda}, {Gavel}, {Gavras}, {Gerssen}, {Geyer}, {Giacobbe},
  {Gilmore}, {Girona}, {Giuffrida}, {Glass}, {Gomes}, {Granvik}, {Gueguen},
  {Guerrier}, {Guiraud}, {Guti{\'e}}, {Haigron}, {Hatzidimitriou}, {Hauser},
  {Haywood}, {Heiter}, {Helmi}, {Heu}, {Hilger}, {Hobbs}, {Hofmann}, {Holland},
  {Huckle}, {Hypki}, {Icardi}, {Jan{\ss}en}, {Jevardat de Fombelle}, {Jonker},
  {Juh{\'a}sz}, {Julbe}, {Karampelas}, {Kewley}, {Klar}, {Kochoska}, {Kohley},
  {Kolenberg}, {Kontizas}, {Kontizas}, {Koposov}, {Kordopatis},
  {Kostrzewa-Rutkowska}, {Koubsky}, {Lambert}, {Lanza}, {Lasne}, {Lavigne}, {Le
  Fustec}, {Le Poncin-Lafitte}, {Lebreton}, {Leccia}, {Leclerc},
  {Lecoeur-Taibi}, {Lenhardt}, {Leroux}, {Liao}, {Licata}, {Lindstr{\o}m},
  {Lister}, {Livanou}, {Lobel}, {L{\'o}pez}, {Managau}, {Mann}, {Mantelet},
  {Marchal}, {Marchant}, {Marconi}, {Marinoni}, {Marschalk{\'o}}, {Marshall},
  {Martino}, {Marton}, {Mary}, {Massari}, {Matijevi{\v{c}}}, {Mazeh},
  {McMillan}, {Messina}, {Michalik}, {Millar}, {Molina}, {Molinaro},
  {Moln{\'a}r}, {Montegriffo}, {Mor}, {Morbidelli}, {Morel}, {Morris},
  {Mulone}, {Muraveva}, {Musella}, {Nelemans}, {Nicastro}, {Noval},
  {O'Mullane}, {Ord{\'e}novic}, {Ord{\'o}{\~n}ez-Blanco}, {Osborne}, {Pagani},
  {Pagano}, {Pailler}, {Palacin}, {Palaversa}, {Panahi}, {Pawlak},
  {Piersimoni}, {Pineau}, {Plachy}, {Plum}, {Poggio}, {Poujoulet},
  {Pr{\v{s}}a}, {Pulone}, {Racero}, {Ragaini}, {Rambaux}, {Ramos-Lerate},
  {Regibo}, {Reyl{\'e}}, {Riclet}, {Ripepi}, {Riva}, {Rivard}, {Rixon},
  {Roegiers}, {Roelens}, {Romero-G{\'o}mez}, {Rowell}, {Royer}, {Ruiz-Dern},
  {Sadowski}, {Sagrist{\`a} Sell{\'e}s}, {Sahlmann}, {Salgado}, {Salguero},
  {Sanna}, {Santana-Ros}, {Sarasso}, {Savietto}, {Schultheis}, {Sciacca},
  {Segol}, {Segovia}, {S{\'e}gransan}, {Shih}, {Siltala}, {Silva}, {Smart},
  {Smith}, {Solano}, {Solitro}, {Sordo}, {Soria Nieto}, {Souchay}, {Spagna},
  {Spoto}, {Stampa}, {Steele}, {Steidelm{\"u}ller}, {Stephenson}, {Stoev},
  {Suess}, {Surdej}, {Szabados}, {Szegedi-Elek}, {Tapiador}, {Taris}, {Tauran},
  {Taylor}, {Teixeira}, {Terrett}, {Teyssandier}, {Thuillot}, {Titarenko},
  {Torra Clotet}, {Turon}, {Ulla}, {Utrilla}, {Uzzi}, {Vaillant}, {Valentini},
  {Valette}, {van Elteren}, {Van Hemelryck}, {Vaschetto}, {Vecchiato},
  {Veljanoski}, {Viala}, {Vicente}, {Vogt}, {von Essen}, {Voss}, {Votruba},
  {Voutsinas}, {Walmsley}, {Weiler}, {Wertz}, {Wevers}, {Wyrzykowski},
  {Yoldas}, {{\v{Z}}erjal}, {Ziaeepour}, {Zorec}, {Zschocke}, {Zucker},
  {Zurbach}, \& {Zwitter}}]{gaiahr2018}
{Gaia Collaboration}, {Babusiaux}, C., {van Leeuwen}, F., {et~al.} 2018, \aap,
  616, A10, \dodoi{10.1051/0004-6361/201832843}

\bibitem[{{Gaia Collaboration} {et~al.}(2021){Gaia Collaboration}, {Brown},
  {Vallenari}, {Prusti}, {de Bruijne}, {Babusiaux}, {Biermann}, {Creevey},
  {Evans}, {Eyer}, {Hutton}, {Jansen}, {Jordi}, {Klioner}, {Lammers},
  {Lindegren}, {Luri}, {Mignard}, {Panem}, {Pourbaix}, {Randich}, {Sartoretti},
  {Soubiran}, {Walton}, {Arenou}, {Bailer-Jones}, {Bastian}, {Cropper},
  {Drimmel}, {Katz}, {Lattanzi}, {van Leeuwen}, {Bakker}, {Cacciari},
  {Casta{\~n}eda}, {De Angeli}, {Ducourant}, {Fabricius}, {Fouesneau},
  {Fr{\'e}mat}, {Guerra}, {Guerrier}, {Guiraud}, {Jean-Antoine Piccolo},
  {Masana}, {Messineo}, {Mowlavi}, {Nicolas}, {Nienartowicz}, {Pailler},
  {Panuzzo}, {Riclet}, {Roux}, {Seabroke}, {Sordo}, {Tanga}, {Th{\'e}venin},
  {Gracia-Abril}, {Portell}, {Teyssier}, {Altmann}, {Andrae}, {Bellas-Velidis},
  {Benson}, {Berthier}, {Blomme}, {Brugaletta}, {Burgess}, {Busso}, {Carry},
  {Cellino}, {Cheek}, {Clementini}, {Damerdji}, {Davidson}, {Delchambre},
  {Dell'Oro}, {Fern{\'a}ndez-Hern{\'a}ndez}, {Galluccio}, {Garc{\'\i}a-Lario},
  {Garcia-Reinaldos}, {Gonz{\'a}lez-N{\'u}{\~n}ez}, {Gosset}, {Haigron},
  {Halbwachs}, {Hambly}, {Harrison}, {Hatzidimitriou}, {Heiter},
  {Hern{\'a}ndez}, {Hestroffer}, {Hodgkin}, {Holl}, {Jan{\ss}en}, {Jevardat de
  Fombelle}, {Jordan}, {Krone-Martins}, {Lanzafame}, {L{\"o}ffler}, {Lorca},
  {Manteiga}, {Marchal}, {Marrese}, {Moitinho}, {Mora}, {Muinonen}, {Osborne},
  {Pancino}, {Pauwels}, {Petit}, {Recio-Blanco}, {Richards}, {Riello},
  {Rimoldini}, {Robin}, {Roegiers}, {Rybizki}, {Sarro}, {Siopis}, {Smith},
  {Sozzetti}, {Ulla}, {Utrilla}, {van Leeuwen}, {van Reeven}, {Abbas}, {Abreu
  Aramburu}, {Accart}, {Aerts}, {Aguado}, {Ajaj}, {Altavilla}, {{\'A}lvarez},
  {{\'A}lvarez Cid-Fuentes}, {Alves}, {Anderson}, {Anglada Varela}, {Antoja},
  {Audard}, {Baines}, {Baker}, {Balaguer-N{\'u}{\~n}ez}, {Balbinot}, {Balog},
  {Barache}, {Barbato}, {Barros}, {Barstow}, {Bartolom{\'e}}, {Bassilana},
  {Bauchet}, {Baudesson-Stella}, {Becciani}, {Bellazzini}, {Bernet}, {Bertone},
  {Bianchi}, {Blanco-Cuaresma}, {Boch}, {Bombrun}, {Bossini}, {Bouquillon},
  {Bragaglia}, {Bramante}, {Breedt}, {Bressan}, {Brouillet}, {Bucciarelli},
  {Burlacu}, {Busonero}, {Butkevich}, {Buzzi}, {Caffau}, {Cancelliere},
  {C{\'a}novas}, {Cantat-Gaudin}, {Carballo}, {Carlucci}, {Carnerero},
  {Carrasco}, {Casamiquela}, {Castellani}, {Castro-Ginard}, {Castro Sampol},
  {Chaoul}, {Charlot}, {Chemin}, {Chiavassa}, {Cioni}, {Comoretto}, {Cooper},
  {Cornez}, {Cowell}, {Crifo}, {Crosta}, {Crowley}, {Dafonte}, {Dapergolas},
  {David}, {David}, {de Laverny}, {De Luise}, {De March}, {De Ridder}, {de
  Souza}, {de Teodoro}, {de Torres}, {del Peloso}, {del Pozo}, {Delbo},
  {Delgado}, {Delgado}, {Delisle}, {Di Matteo}, {Diakite}, {Diener},
  {Distefano}, {Dolding}, {Eappachen}, {Edvardsson}, {Enke}, {Esquej}, {Fabre},
  {Fabrizio}, {Faigler}, {Fedorets}, {Fernique}, {Fienga}, {Figueras},
  {Fouron}, {Fragkoudi}, {Fraile}, {Franke}, {Gai}, {Garabato},
  {Garcia-Gutierrez}, {Garc{\'\i}a-Torres}, {Garofalo}, {Gavras}, {Gerlach},
  {Geyer}, {Giacobbe}, {Gilmore}, {Girona}, {Giuffrida}, {Gomel}, {Gomez},
  {Gonzalez-Santamaria}, {Gonz{\'a}lez-Vidal}, {Granvik},
  {Guti{\'e}rrez-S{\'a}nchez}, {Guy}, {Hauser}, {Haywood}, {Helmi}, {Hidalgo},
  {Hilger}, {H{\l}adczuk}, {Hobbs}, {Holland}, {Huckle}, {Jasniewicz},
  {Jonker}, {Juaristi Campillo}, {Julbe}, {Karbevska}, {Kervella}, {Khanna},
  {Kochoska}, {Kontizas}, {Kordopatis}, {Korn}, {Kostrzewa-Rutkowska},
  {Kruszy{\'n}ska}, {Lambert}, {Lanza}, {Lasne}, {Le Campion}, {Le Fustec},
  {Lebreton}, {Lebzelter}, {Leccia}, {Leclerc}, {Lecoeur-Taibi}, {Liao},
  {Licata}, {Lindstr{\o}m}, {Lister}, {Livanou}, {Lobel}, {Madrero Pardo},
  {Managau}, {Mann}, {Marchant}, {Marconi}, {Marcos Santos}, {Marinoni},
  {Marocco}, {Marshall}, {Martin Polo}, {Mart{\'\i}n-Fleitas}, {Masip},
  {Massari}, {Mastrobuono-Battisti}, {Mazeh}, {McMillan}, {Messina},
  {Michalik}, {Millar}, {Mints}, {Molina}, {Molinaro}, {Moln{\'a}r},
  {Montegriffo}, {Mor}, {Morbidelli}, {Morel}, {Morris}, {Mulone}, {Munoz},
  {Muraveva}, {Murphy}, {Musella}, {Noval}, {Ord{\'e}novic}, {Orr{\`u}},
  {Osinde}, {Pagani}, {Pagano}, {Palaversa}, {Palicio}, {Panahi}, {Pawlak},
  {Pe{\~n}alosa Esteller}, {Penttil{\"a}}, {Piersimoni}, {Pineau}, {Plachy},
  {Plum}, {Poggio}, {Poretti}, {Poujoulet}, {Pr{\v{s}}a}, {Pulone}, {Racero},
  {Ragaini}, {Rainer}, {Raiteri}, {Rambaux}, {Ramos}, {Ramos-Lerate}, {Re
  Fiorentin}, {Regibo}, {Reyl{\'e}}, {Ripepi}, {Riva}, {Rixon}, {Robichon},
  {Robin}, {Roelens}, {Rohrbasser}, {Romero-G{\'o}mez}, {Rowell}, {Royer},
  {Rybicki}, {Sadowski}, {Sagrist{\`a} Sell{\'e}s}, {Sahlmann}, {Salgado},
  {Salguero}, {Samaras}, {Sanchez Gimenez}, {Sanna}, {Santove{\~n}a},
  {Sarasso}, {Schultheis}, {Sciacca}, {Segol}, {Segovia}, {S{\'e}gransan},
  {Semeux}, {Shahaf}, {Siddiqui}, {Siebert}, {Siltala}, {Slezak}, {Smart},
  {Solano}, {Solitro}, {Souami}, {Souchay}, {Spagna}, {Spoto}, {Steele},
  {Steidelm{\"u}ller}, {Stephenson}, {S{\"u}veges}, {Szabados}, {Szegedi-Elek},
  {Taris}, {Tauran}, {Taylor}, {Teixeira}, {Thuillot}, {Tonello}, {Torra},
  {Torra}, {Turon}, {Unger}, {Vaillant}, {van Dillen}, {Vanel}, {Vecchiato},
  {Viala}, {Vicente}, {Voutsinas}, {Weiler}, {Wevers}, {Wyrzykowski}, {Yoldas},
  {Yvard}, {Zhao}, {Zorec}, {Zucker}, {Zurbach}, \& {Zwitter}}]{Gaia2021}
{Gaia Collaboration}, {Brown}, A.~G.~A., {Vallenari}, A., {et~al.} 2021, \aap,
  649, A1, \dodoi{10.1051/0004-6361/202039657}

\bibitem[{{Goodwin}(2013)}]{2013MNRAS.430L...6G}
{Goodwin}, S.~P. 2013, \mnras, 430, L6, \dodoi{10.1093/mnrasl/sls037}

\bibitem[{{Gullikson} {et~al.}(2016){Gullikson}, {Kraus}, \&
  {Dodson-Robinson}}]{gullikson2016}
{Gullikson}, K., {Kraus}, A., \& {Dodson-Robinson}, S. 2016, \aj, 152, 40,
  \dodoi{10.3847/0004-6256/152/2/40}

\bibitem[{{Henry} \& {McCarthy}(1993)}]{1993AJ....106..773H}
{Henry}, T.~J., \& {McCarthy}, Donald~W., J. 1993, \aj, 106, 773,
  \dodoi{10.1086/116685}

\bibitem[{{Hwang} {et~al.}(2022){Hwang}, {Ting}, \& {Zakamska}}]{Hwang2022}
{Hwang}, H.-C., {Ting}, Y.-S., \& {Zakamska}, N.~L. 2022, \mnras, 512, 3383,
  \dodoi{10.1093/mnras/stac675}

\bibitem[{{Ivezi{\'c}} {et~al.}(2019){Ivezi{\'c}}, {Connelly}, {Vanderplas}, \&
  {Gray}}]{ivezic2019}
{Ivezi{\'c}}, {\v{Z}}., {Connelly}, A.~J., {Vanderplas}, J.~T., \& {Gray}, A.
  2019, {Statistics, Data Mining, and Machine Learning in Astronomy}

\bibitem[{{Kl{\"u}ter} {et~al.}(2020){Kl{\"u}ter}, {Bastian}, \&
  {Wambsganss}}]{Kluter202}
{Kl{\"u}ter}, J., {Bastian}, U., \& {Wambsganss}, J. 2020, \aap, 640, A83,
  \dodoi{10.1051/0004-6361/201937061}

\bibitem[{{Kroupa} {et~al.}(1993){Kroupa}, {Tout}, \& {Gilmore}}]{kroupa93}
{Kroupa}, P., {Tout}, C.~A., \& {Gilmore}, G. 1993, \mnras, 262, 545,
  \dodoi{10.1093/mnras/262.3.545}

\bibitem[{{Lucy}(2014)}]{lucy2014}
{Lucy}, L.~B. 2014, \aap, 563, A126, \dodoi{10.1051/0004-6361/201322649}

\bibitem[{Mann {et~al.}(2019)Mann, Dupuy, Kraus, Gaidos, Ansdell, Ireland,
  Rizzuto, Hung, Dittmann, Factor, Feiden, Martinez,
  Ru{\'{\i}}z-Rodr{\'{\i}}guez, \& Thao}]{Mann_2019}
Mann, A.~W., Dupuy, T., Kraus, A.~L., {et~al.} 2019, The Astrophysical Journal,
  871, 63, \dodoi{10.3847/1538-4357/aaf3bc}

\bibitem[{{Moe} \& {Di Stefano}(2017)}]{moe2017}
{Moe}, M., \& {Di Stefano}, R. 2017, \apjs, 230, 15,
  \dodoi{10.3847/1538-4365/aa6fb6}

\bibitem[{Pearce {et~al.}(2020)Pearce, Kraus, Dupuy, Mann, Newton, Tofflemire,
  \& Vanderburg}]{Pearce_2020}
Pearce, L.~A., Kraus, A.~L., Dupuy, T.~J., {et~al.} 2020, The Astrophysical
  Journal, 894, 115, \dodoi{10.3847/1538-4357/ab8389}

\bibitem[{{Pittordis} \& {Sutherland}(2019)}]{2019MNRAS.488.4740P}
{Pittordis}, C., \& {Sutherland}, W. 2019, \mnras, 488, 4740,
  \dodoi{10.1093/mnras/stz1898}

\bibitem[{{Pittordis} \& {Sutherland}(2022)}]{2022arXiv220502846P}
---. 2022, arXiv e-prints, arXiv:2205.02846.
\newblock \doarXiv{2205.02846}

\bibitem[{{Raghavan} {et~al.}(2010){Raghavan}, {McAlister}, {Henry}, {Latham},
  {Marcy}, {Mason}, {Gies}, {White}, \& {ten Brummelaar}}]{Raghavan2010}
{Raghavan}, D., {McAlister}, H.~A., {Henry}, T.~J., {et~al.} 2010, \apjs, 190,
  1, \dodoi{10.1088/0067-0049/190/1/1}

\bibitem[{{Rana}(1987)}]{rana1987}
{Rana}, N.~C. 1987, \aap, 184, 104

\bibitem[{{Reyl{\'e}}(2018)}]{rayle2018}
{Reyl{\'e}}, C. 2018, \aap, 619, L8, \dodoi{10.1051/0004-6361/201834082}

\bibitem[{{Serenelli} {et~al.}(2021){Serenelli}, {Weiss}, {Aerts}, {Angelou},
  {Baroch}, {Bastian}, {Beck}, {Bergemann}, {Bestenlehner}, {Czekala},
  {Elias-Rosa}, {Escorza}, {Van Eylen}, {Feuillet}, {Gandolfi}, {Gieles},
  {Girardi}, {Lebreton}, {Lodieu}, {Martig}, {Miller Bertolami}, {Mombarg},
  {Morales}, {Moya}, {Nsamba}, {Pavlovski}, {Pedersen}, {Ribas}, {Schneider},
  {Silva Aguirre}, {Stassun}, {Tolstoy}, {Tremblay}, \&
  {Zwintz}}]{Serenelli2021}
{Serenelli}, A., {Weiss}, A., {Aerts}, C., {et~al.} 2021, \aapr, 29, 4,
  \dodoi{10.1007/s00159-021-00132-9}

\bibitem[{{Shatsky}(2001)}]{shatsky2001}
{Shatsky}, N. 2001, \aap, 380, 238, \dodoi{10.1051/0004-6361:20011401}

\bibitem[{{Sollima}(2019)}]{sollima2019}
{Sollima}, A. 2019, \mnras, 489, 2377, \dodoi{10.1093/mnras/stz2093}

\bibitem[{{Southworth}(2015)}]{Southworth2015}
{Southworth}, J. 2015, in Astronomical Society of the Pacific Conference
  Series, Vol. 496, Living Together: Planets, Host Stars and Binaries, ed.
  S.~M. {Rucinski}, G.~{Torres}, \& M.~{Zejda}, 164.
\newblock \doarXiv{1411.1219}

\bibitem[{{Stassun} {et~al.}(2018){Stassun}, {Corsaro}, {Pepper}, \&
  {Gaudi}}]{Stassun2018}
{Stassun}, K.~G., {Corsaro}, E., {Pepper}, J.~A., \& {Gaudi}, B.~S. 2018, \aj,
  155, 22, \dodoi{10.3847/1538-3881/aa998a}

\bibitem[{{Tokovinin}(2014)}]{tokovinin2014}
{Tokovinin}, A. 2014, \aj, 147, 87, \dodoi{10.1088/0004-6256/147/4/87}

\bibitem[{{Tokovinin} \& {Kiyaeva}(2016)}]{Tokovinin2016}
{Tokovinin}, A., \& {Kiyaeva}, O. 2016, \mnras, 456, 2070,
  \dodoi{10.1093/mnras/stv2825}

\bibitem[{{Tokovinin}(1998)}]{tokovinin1998}
{Tokovinin}, A.~A. 1998, Astronomy Letters, 24, 178

\bibitem[{{Torres}(2019)}]{80Tau}
{Torres}, G. 2019, \apj, 883, 105, \dodoi{10.3847/1538-4357/ab3a30}

\bibitem[{{Wittenmyer} {et~al.}(2020){Wittenmyer}, {Wang}, {Horner}, {Butler},
  {Tinney}, {Carter}, {Wright}, {Jones}, {Bailey}, {O'Toole}, \&
  {Johns}}]{wittenmyer2020}
{Wittenmyer}, R.~A., {Wang}, S., {Horner}, J., {et~al.} 2020, \mnras, 492, 377,
  \dodoi{10.1093/mnras/stz3436}

\bibitem[{{Ziegler} {et~al.}(2018){Ziegler}, {Law}, {Baranec}, {Morton},
  {Riddle}, {De Lee}, {Huber}, {Mahadevan}, \& {Pepper}}]{zeigler2018}
{Ziegler}, C., {Law}, N.~M., {Baranec}, C., {et~al.} 2018, \aj, 156, 259,
  \dodoi{10.3847/1538-3881/aad80a}

\end{thebibliography}
\bibliographystyle{aasjournal}



\end{document}